\documentclass[times, twoside]{StyleBioRxiv}
\usepackage{blindtext}
\leadauthor{} 

\begin{document}

\title{Hierarchical defect-induced condensation in active nematics}
\shorttitle{}

\author[a,*]{Timo Kr\"uger}

\author[a,*]{Ivan Maryshev}

\author[a,b,1]{Erwin Frey}
\affil[a]{Arnold Sommerfeld Center for Theoretical Physics (ASC) and Center for NanoScience (CeNS), Department of Physics, Ludwig-Maximilians-Universität München,
Theresienstrasse 37, 80333 Munich, Germany}

\affil[b]{Max Planck School Matter to Life, Hofgartenstraße 8, 80539 Munich, Germany}
\affil[*]{T.K. and I.M. contributed equally to this work.} 
\affil[1]{Corresponding author: frey@lmu.de}

\maketitle

\begin{abstract}
Topological defects play a central role in the formation and organization of various biological systems.
Historically, such nonequilibrium defects have been mainly studied in the context of homogeneous active nematics. 
Phase-separated systems, in turn, are known to form dense and dynamic nematic bands, but typically lack topological defects.
In this paper, we use agent-based simulations of weakly aligning, self-propelled polymers and demonstrate that contrary to the existing paradigm phase-separated active nematics form ${-1/2}$ defects. Moreover, these defects, emerging due to interactions among dense nematic bands, constitute a novel second-order collective state. We investigate the morphology of defects in detail and find that their cores correspond to a strong increase in density, associated with a condensation of nematic fluxes. Unlike their analogs in homogeneous systems, such condensed defects form and decay in a different way and do not involve positively charged partners.
We additionally observe and characterize lateral arc-like structures that separate from a band's bulk and move in transverse direction.
We show that the key control parameters defining the route from stable bands to the coexistence of dynamic lanes and defects are the total density of particles and their path persistence length.
We introduce a hydrodynamic theory that qualitatively recapitulates all the main features of the agent-based model, and use it to show that the emergence of both defects and arcs can be attributed to the same anisotropic active fluxes.
Finally, we present a way to artificially engineer and position defects, and speculate about experimental verification of the provided model.
\end {abstract}

\section*{Introduction}
The characteristic features of a nematic liquid crystal are the emergence of long-range orientational order and the occurrence of \textit{half-integer topological defects}, which, however, are annealed at thermodynamic equilibrium~\cite{de1993physics}. 
The dynamics of its nonequilibrium counterpart, an active nematic~\cite{Marchetti2013, Doostmohammadi2018}, is in contrast governed by the persistent creation and annihilation of pairs of topological defects with opposite charges, leading to a dynamic steady state commonly referred to as active turbulence \cite{alert2021active}.
Dense gel-like mixtures of microtubules (cytoskeletal filaments) and kinesins (molecular motors) that cause relative sliding between microtubules have become experimental platforms for studying the formation, dynamics, and annihilation of these toplogical defects~\cite{sanchez_spontaneous_2012,Decamp2015}. 
The observed complex defect dynamics have been investigated using hydrodynamic theories~\cite{giomi_defect_2013,shankar_defect_2018}.
The basic insight derived from such studies is that topological defects constantly generate active flow in momentum-conserving systems \cite{thampi_instabilities_2014,Giomi2014} or active flux in momentum non-conserving systems \cite{putzig_instabilities_2016,Maryshev2019Dry}.

Another experimental model system for active nematics is the actomyosin motility assay, in which actin filaments actively glide over a lawn of myosin motor proteins, performing a persistent random walk with constant speed~\cite{schaller_polar_2010,butt_myosin_2010}.
These systems exhibit \textit{phase separation} into dense polar-ordered regions and dilute disordered regions, which is further corroborated by numerical analyses of corresponding theoretical models~\cite{gregoire_onset_2004,solon_phase_2015,huber_microphase_2021}. 
Tuning the interaction between actin filaments by the addition of polyethylene glycol led to the emergence of a dynamic coexistence of ordered states with fluctuating nematic and polar symmetry~\cite{Huber2018}, which has been explained by pattern-induced symmetry breaking~\cite{denk_pattern-induced_2020-1}. Systems exhibiting dense, purely nematic lanes have been thoroughly investigated by both simulations and hydrodynamic theories \cite{ginelli_large-scale_2010,
Peshkov2012,
ngo_large-scale_2014,
grosmann_mesoscale_2016,
Maryshev2019Dry,
Maryshev2020,
Cai2019,
grosmann_particle-field_2020}.

As for half-integer topological defects, the common paradigm states that they are absent in dilute self-propelled active nematics \cite{chate_dry_2020}, but fundamental exclusion criteria for their existence have not been given.
In fact, no steady-state topological defects have yet been found in this subclass of strongly phase-separated active matter.
So far, it has only been observed that transient defects can occur in models with weak density inhomogeneity during the coarsening process \cite{mishra2014aspects}.
Moreover, toy models inspired by dilute nematic systems \textit{without} self-propulsion can exhibit defect formation \cite{Maryshev2020}. 
However, the authors attest that the connection of their phenomenological theory to existing experimental systems is tenuous.

Here we investigate dilute active nematics for the presence of defects using an agent-based model of ``weakly-aligning self-propelled polymers'' (WASP) which has been shown to faithfully reproduce the behavior of real actomyosin motility assays on all relevant length and timescales including pattern formation processes and the topology of the phase diagram~\cite{Huber2018,huber_microphase_2021}.
This allows us to leverage these agent-based simulations as an \textit{in-silico} experimental system with which to discover new phenomena. 
We show that the two hitherto seemingly incompatible phenomena --- phase separation and topological defects --- are actually closely linked in weakly interacting active nematics.

In particular, we characterize a subclass of topological defects associated with the compression of nematic fluxes, which are similar to phenomena predicted in conceptual models \cite{mishra2014aspects,Maryshev2020}, albeit in a different context.
These defects appear as characteristic collective excitations in a novel nonequilibrium steady state. They are in dynamic equilibrium with nematic lanes from which they emerge and into which they disassemble.
Additionally, we find another type of topologically charged structure, filamentous arc ejections (FAEs) --- elongated arc-shaped  polymer bundles that detach from nematic bands --- remotely resembling $+1/2$ defects.
To elucidate the mechanisms underlying these phenomena, we also introduce a hydrodynamic theory, building on previously published models~\cite{Maryshev2019Dry,Maryshev2020}. 
Exploiting the respective strengths of these two complementary theoretical approaches, we uncover a close relationship between the dynamics of phase-separated nematic bands, formation of topologically charged structures, and the associated condensation phenomena.

\section*{Results}
\subsection*{Simulation setup}
\begin{figure}[!t]
\centering
\includegraphics[width=\linewidth]{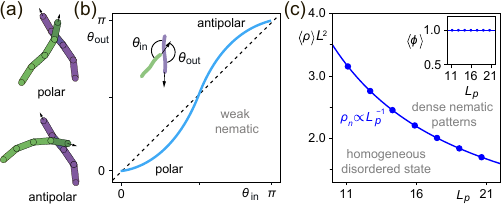}
\caption{
Nematic interaction between polymers and onset of order. (a) Schematic depiction of two interacting polymers. Depending on whether or not the impact angle is smaller or larger $\frac{\pi}{2}$, polymer directions are either aligned (\textit{upper panel}) or anti-aligned (\textit{lower panel}). (b) 
Illustration of the binary collision statistics corresponding to a weak nematic interaction between polymers, which is symmetric with respect to the point $(\frac{\pi}{2},\frac{\pi}{2})$. (c)  
Phase diagram of active nematics with collision statistics shown in panel (b). 
The blue line shows the density corresponding to isotropic-nematic transition $\rho_n$, which is inversely proportional to the persistence length
$\rho_n \propto L_p ^{-1}$. Please refer to 
the SI for details. 
\textit{Inset}: The same graph plotted as a function of $\langle \phi \rangle = \langle \rho\rangle/\rho_n$ and $L_p$. 
}
\label{fig:fig1}
\end{figure}

We use agent-based simulations that emulate the dynamics of \textit{weakly interacting self-propelled polymers} (WASP) of fixed length $L$ on two-dimensional surfaces building on earlier work~\cite{Huber2018, huber_microphase_2021}; refer to the SI for further details on the algorithm. 
Each polymer consists of a tail pulled by a tip that follows a trajectory corresponding to a persistent random walk with persistence length $L_p$.
Upon collision of a polymer tip with the contour of another polymer, a weak alignment torque is assumed to act that changes its direction of motion [Fig.~\ref{fig:fig1}(a)]. 
Here we use a purely nematic alignment interaction [Fig.~\ref{fig:fig1}(b)] whose strength is set by the parameter $\alpha_n$. 
Additionally, a small repulsion force $F$ acts on polymer tips that overlap with other polymers. 

Here we are interested in systems that have a collision statistics with \textit{purely nematic symmetry} [Fig.~\ref{fig:fig1}(b)].
Figure~\ref{fig:fig1}(c) shows the phase diagram of such a weak nematic as a function of the average polymer density $\langle \rho \rangle L^2$ 
and path persistence length $L_p$; hereafter $\langle ... \rangle$ denotes spatial averaging.
It exhibits an isotropic-nematic transition from a disordered homogeneous phase to a nematically ordered phase. 

The phase boundary $\rho_n (L_p)$ approximately scales as $L_p^{-1}$; refer to 
the SI for details.
Thus, when the phase diagram is redrawn as a function of $L_p$ and the spatially averaged normalized density ${\langle \phi \rangle = \langle \rho \rangle / \rho_n}$, the phase boundary essentially becomes a horizontal line [inset of Fig.~\ref{fig:fig1}(c)].

\subsection*{Dense topologically charged structures}

As expected for nematically interacting systems, our simulations show isolated \textit{nematic lanes} that exhibit strong bending fluctuations on large length and time scales  (cf. Movie S1 SI) caused by lateral instabilities \cite{ginelli_large-scale_2010,bertin_mesoscopic_2013}.

\begin{figure*}[!t]
\centering
\includegraphics[width=0.95\linewidth]{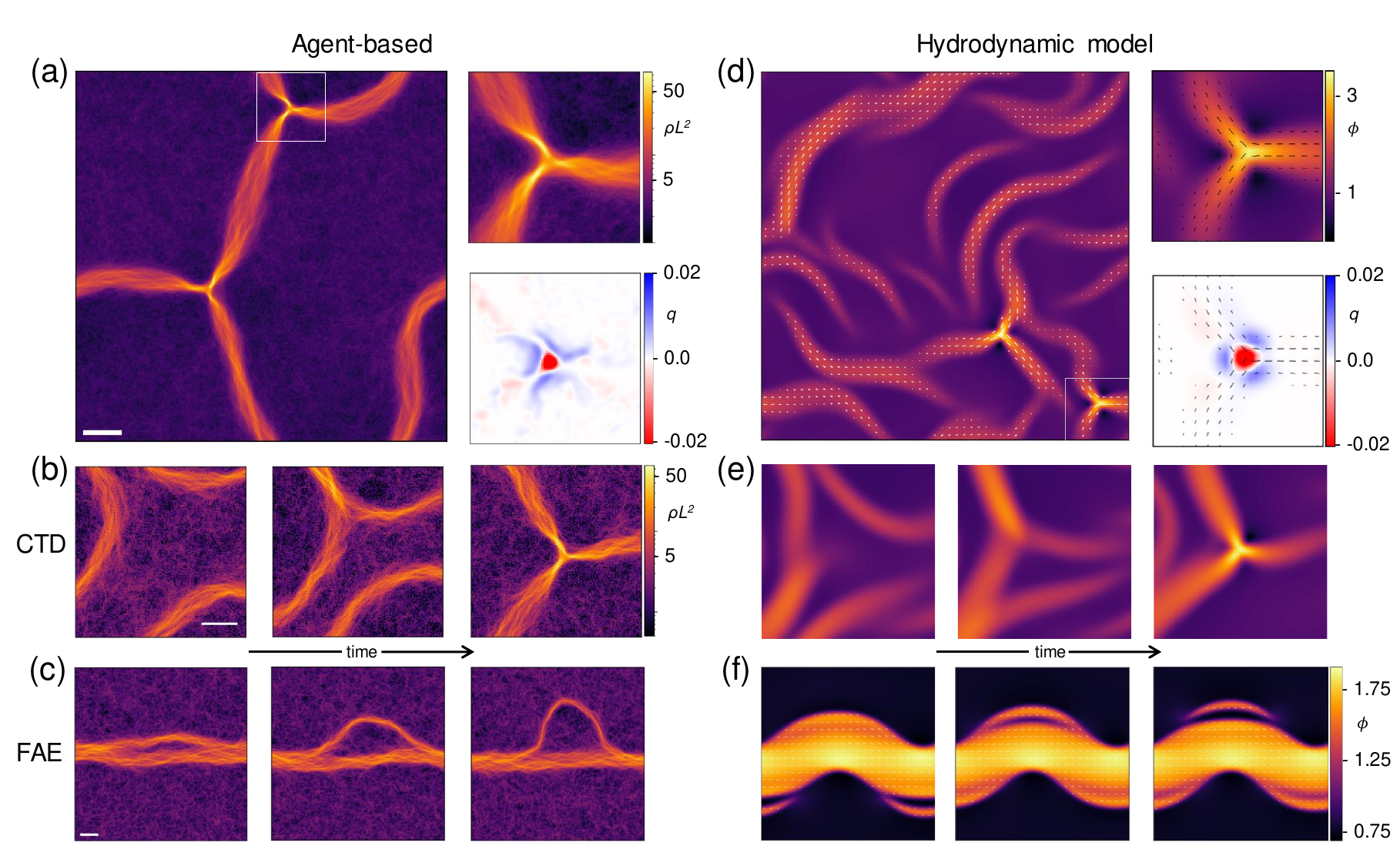}
\caption{Condensed defects and filamentous arc ejections. Left column [(a) to (c)] shows results for agent-based simulations, right column [(d) to (f)] for the hydrodynamic model. [(a) and (d), \textit{left panels}] Spatial density distribution of a system simultaneously exhibiting two condensed defects. A magnified view of one defect (rectangular marked region) is shown in the \textit{upper right panels} in (a) and (d). The \textit{lower right panels} in (a) and (d) show the topological charge density $q$ of the magnified region. In both cases, a $-\frac{1}{2}$ defect is surrounded by positively charged regions of space. 
[(b) and (e)]
Magnified views of the formation of a condensed defect. Three convex bands meet and self-focus to form a dense structure, in the center of which the topological charge that was previously on the outside of the bands is trapped. [(e): same color bar as (d)]
[(c) and (f)] Snapshots of the evolution of a filamentous arc ejection as observed in the agent-based and hydrodynamic models, respectively. [(c): same color bar as (b)]
[Parameters are (a) $\langle \rho \rangle L^2{=}3.5$, $L_p{=}11.1$, (b-c) $\langle \rho \rangle L^2{=}2.7$, $L_p{=}14.3$, (a-c) Scale bars: 15L; see Appendix
for further parameters].
} 
\label{fig:fig2}
\end{figure*}

In our simulations, in addition to these typical nematic lanes, we also discover distinct types of  topologically charged structures. 
One class of these are three-armed filamentous structures containing a topological defect with charge $-1/2$ at their center [Fig.~\ref{fig:fig2}(a)]. 
They are typically formed when three curved nematic lanes --- with their convex sides facing each other --- meet and condense into a topological defect with a high-density core region [Fig.~\ref{fig:fig2}(b)]; we do not observe ``collisions'' of four lanes.
Unlike defects in non phase-separated active nematics, these \textit{condensed topological defects} (CTDs) do not have a directly corresponding positively charged partner. 
Instead, they are surrounded by an extended topologically charged region with a dispersed positive charge, as can be seen in Fig.~\ref{fig:fig2}(a) (lower right panel), which depicts the topological charge density as defined in Refs.~\cite{Blow2014,Maryshev2020}.
Moreover, our simulations show that the active nematic flux is gradually compressed as the triple junction of the nematic lanes (defect core) is approached
[Fig.~\ref{fig:fig2}(a), top right panel]. 
This leads to a reduction in lane width and a corresponding increase in density, which reaches a maximum in proximity of the core.
These three-armed topological defects are dynamic structures that are constantly being dissolved and reassembled. 
A second class of structures we observe are lateral filamentous arcs that separate from the bulk of a straight nematic band and eventually move in transverse direction.
A time trace of such a \textit{filamentous arc ejection} (FAE) is shown in Fig.~\ref{fig:fig2}(c).
These structures have similarities to $+1/2$ defects: they are ``curved'' and they always emanate in the direction of their convex side. 
Somewhat similar observations have been made in continuum models constructed for nematic particles with velocity reversals~\cite{ngo_large-scale_2014}. However, the authors did not address the properties of these structures or the reasons underlying their formation. 
While there are certainly similarities on a superficial phenomenological level between FAEs and these structures, the underlying mechanisms and nature of these structures may be quite different.

Having discovered these collective topological structures in our \textit{in-silico} experiments, we sought to explore how their emergence is affected by a change of parameters. 
However, since the lateral instabilities of nematic bands required for the formation of CTDs (cf. section ``From CTDs to FAEs and bands'' below) occur only on very long time scales, a systematic investigation of a phase diagram in agent-based simulation is numerically prohibitively demanding.
Therefore, we sought an alternative way to explore the spatiotemporal dynamics of the systems that would enable us to dissect the processes underlying the formation of CTDs and FAEs. 
As explained next, we achieved this through constructing a hydrodynamic approach that captures all the main features of our agent-based simulation setup.

\subsection*{Hydrodynamic model provides access to the phase diagram}
To this end we used the standard Boltzmann-like approach 
(see SI).
However, as discussed below, this model was insufficient to explain the emergence of half-integer defects and was therefore generalized to include density-dependent corrections.

By analogy with passive \textit{model C} in the Hohenberg-Halperin classification scheme~\cite{HohenbergHalperin} we formulate a hydrodynamic model in terms of a density and an order parameter field. 
For an active nematic, these are the (normalized) polymer density 
${\phi = \int\text d \theta  \, P(\theta)/ \rho_n}$,
and the traceless and symmetric tensor 
${Q_{ij} = \int\text d \theta \, P(\theta)(2n_in_j- \delta_{ij})}$ (nematic order parameter), where the unit vector ${\mathbf n= (n_x,n_y)=(\text {cos} \,\theta, \text {sin}\, \theta)}$ defines the the local polymer orientation vector and $P(\theta)$ denotes the probability density for the polymer orientation $\theta$.
The eigenvector associated with the larger of the two eigenvalues of the $Q$-tensor can be viewed as depicting the average orientation of the polymers.

Unlike classical \textit{model C}, however, a hydrodynamic model for active nematics must be intrinsically nonequilibrium in character and its dynamics can not be determined by the gradient descent in a single free-energy landscape.
Nevertheless, using the analogy to the dynamics near thermal equilibrium, some intuition can be gained for the design of the model.
As we discuss in more detail below, part of the system's dynamics can be understood in terms of two separate effective free-energy functionals for the non-conservative $Q$-tensor ($\cal{F_{Q}}$) and the conservative density field ($\cal{F_{\phi}}$), similar to related nonequilibrium models discussed recently \cite{li_hierarchical_2021}.

Mass-conservation requires that the density obeys a continuity equation $\partial_t \phi = - \partial_i J_i$.
In general, for symmetry reasons, the current must be the gradient of a scalar quantity and a tensorial quantity containing the $Q$-tensor. 
Similar to model B, the scalar component is of the form 
$ J_i^{\text{iso}} = -\partial_i \, \mu (\phi) $ 
with chemical potential 
$ \mu (\phi) = \nu (\phi) \, \phi $.
Here, the first and second terms of ${\nu (\phi) = \lambda^2+\nu_{\phi}\phi}$ account for motility-induced  effective diffusion with the diffusion constant $\lambda^2 \propto L_p^2$ \cite{baskaran_self-regulation_2012}, and for steric repulsion due to excluded-volume interactions~\cite{ahmadi2006hydrodynamics, baskaran2010nonequilibrium, Maryshev2018}, respectively. The latter contribution represents the density-dependent correction.

For the tensorial part, we write ${J_i^{\text{aniso}} \,{=}\, {-} \partial_j  \big[\chi(\phi) Q_{ij}\big]}$, which again is assumed to contain motility- and interaction-induced parts: $\chi (\phi) =\lambda^2+\chi_{\phi}\phi$. Similar as above, the latter term represents the density-dependent correction motivated by theories for active nematics \cite{Maryshev2019Dry,Cai2019}, and it is controlled by the phenomenological parameter $\chi_{\phi}$.
It will turn out that this anisotropic term leads to phase separation, since it causes compression in the direction  perpendicular to the axis of the local orientational order.
Taken together, one gets
\begin{align}
    \partial_{t} \phi
    =
    \partial_{i}\partial_{j}
    \big[
    \nu(\phi)\phi \, 
    \delta_{ij}
    +
    \chi(\phi) \, Q_{i j}
    \big] 
    \, .
    \label{eqrho}
\end{align}
The isotropic flux (first term) can be written in terms of an effective free-energy functional $\cal {F}_{\phi}=$ $\int \mathrm{d}^2 x \, \left(\frac 12 \lambda^2 \phi^{2}+\frac13 \nu_{\phi} \phi^{3}\right)$.  
In contrast, however, the anisotropic flux (second term in \eqref{eqrho}) violates time-reversal symmetry~\cite{cates2019active, shaebani2020computational}.

We assume the time evolution of the nematic tensor to be of the form
\begin{align}
    \partial_{t} Q_{i j}
    = -\Big[
    \frac{\delta {\cal F_{Q}}}{\delta Q_{ij}}\Big]^\text{st}
    =
    -\Big[
    \frac{\delta {\cal F_{Q}}}{\delta Q_{ij}}
    - \frac12 \, 
    \delta_{ij} \, 
    \text{Tr} 
    \Big(\frac{\delta {\cal F_{Q}}}{\delta Q_{ij}}
    \Big)
    \Big]
    , 
\label{eqQ}
\end{align}
which corresponds to a gradient dynamics (model A) determined by the effective free-energy functional ${\cal F_{Q}}$; here and in the following $[...]^\text{st}$ denotes the traceless and symmetric part of a tensor.
We have chosen the timescale such that the friction coefficient in the gradient dynamics is set to $1$.

The effective free-energy functional has a standard \textit{Landau\--deGennes} (LdG) part~\cite{de1993physics} responsible for an isotropic to nematic transition, but also includes a coupling between density gradients and the orientation of polymers as in inhomogeneous active nematics \cite{Cai2019,Maryshev2019Dry},
\begin{align}
    {\cal F_Q}
    =
    \int \! d^2 x
    \Big(
    &\tfrac12
    \big[
    (1-\phi)Q^2
    + \tfrac12\beta (Q^2)^2
    + \kappa \, (\partial_jQ_{ij})^2
    \big]
    \nonumber\\
    &- 
    Q_{ij}\big[
    \omega \, 
    \partial_i\partial_j\phi
    +
    \omega^a(\partial_i\phi)(\partial_j\phi)
    \big]
    \Big)
    \, .
    \label{eq:FQ}
\end{align}

The LdG free-energy density in terms of the order parameter ${Q^2= Q_{kl}Q_{kl}}$ describes a nematic ordering transition at the critical density ${\phi_c = 1}$ with the gradient term playing the role of a generalised elasticity.
The stiffness coefficient (or Frank constant) $\kappa$ also contains two contributions, one from the motility of the polymers \cite{ginelli_large-scale_2010,ngo_large-scale_2014}, and the other due to interactions \cite{Maryshev2019Dry}: ${\kappa (\phi) = \tfrac12 \lambda^2+\kappa_{\phi}\langle\phi\rangle}$.
Note that the last term --- the density-dependent correction to elasticity --- is linearised around the mean value of density $\langle\phi\rangle$
(see SI). 
The second line in \eqref{eq:FQ} takes into account the coupling between density gradients and nematic order, and can be derived solely on the basis of symmetry considerations. 
The functional derivatives of $\cal {F_Q}$ with respect to the nematic tensor correspond to ``interfacial torques'' \cite{Maryshev2019Dry} in the equation of motion for the nematic tensor.
They rotate the director at the interface between high- and low-density domains, where the gradients of $\phi$ are the strongest.

The lowest-order coupling --- and the  associated ``aligning torque'' \cite{Marchetti2013} $\omega \, [\partial_i\partial_j\phi]^\text{st}$ --- 
is iconic for active nematics \cite{Marchetti2013,cates2019active,baskaran_self-regulation_2012,chate_dry_2020}. 
It is responsible for the destabilization of straight nematic lanes, eventually resulting in lane undulations (or other types of chaotic behavior associated with ``dry active turbulence'' \cite{ngo_large-scale_2014,alert2021active,Maryshev2019Dry}). 
In our case, this term is due to self-advection (${\omega=\lambda^2}$, see 
SI)
but it can be considered as ``diffusive'' since anisotropic diffusion of particles leads to an analogous contribution.

Interaction between the polymers yields the next-order couplings in \eqref{eq:FQ}. 
On symmetry grounds there are two different terms quadratic in $\phi$:
$[\phi \, \partial_i\partial_j\phi]^{st}$ and $[(\partial_i\phi) (\partial_j\phi)]^{st}$; both can also be obtained by explicitly coarse-graining microscopic models for interacting active polymers \cite{Maryshev2019Dry}.
The former recalls the diffusive $\omega$-term (especially after the linearization around $\langle\phi\rangle$) and therefore is ignored here.
The latter is associated with torque, which is bilinear in the density gradients $\omega^a [(\partial_i\phi) (\partial_j\phi)]^{st}$, providing an effective liquid-crystalline ``anchoring'' \cite{Maryshev2020, sulaiman2006lattice, araki2004nematohydrodynamic} (or preferred orientation) of the nematic director field with respect to the density gradients.
The parameter $\omega^a$ is taken to be negative to ensure tangential anchoring, implying that polymers tend to orient  perpendicular to the density gradients (or parallel to the boundary of dense lanes). 

For simplicity, we ignore additional non-linearities in the equation of motion for the Q-tensor. Such contributions are considered elsewhere \cite{mishra2010dynamic,Putzig2014,Maryshev2019Dry} where they are typically regarded as a modification to the elasticity terms.

Taken together Eqs.~(\ref{eqrho},~\ref{eqQ}) are a generalization of the \textit{active model C} \cite{Maryshev2019Dry,Maryshev2020}, which was originally introduced for \textit{non self-propelled} biofilaments in the presence of molecular motors. The major difference is that the model now explicitly includes self-propulsion. Moreover, by including density-dependent terms, it shows the same results as the agent-based simulations (see discussion below) and is therefore quantitatively linked to the actomyosin motility assay. Finally, it possesses less degrees of freedom, since most of the terms are rigorously derived and are controlled by the same parameter ($\lambda$).

We consider
$\nu_{\phi},\,\chi_{\phi},\,\kappa_{\phi},\,\omega$ and $\omega^a$ as phenomenological parameters and solve the equations of motion numerically.
This model robustly reproduces the results obtained in the agent-based simulation to a very high degree of fidelity and for a large range of parameters. 
It exhibits CTDs and FAEs whose structure, topological charge, and formation process are very similar to the ones observed in WASP; cf. Fig.~\ref{fig:fig2}(d)-(f). Therefore, in the following we use this hydrodynamic approach to analyse and underpin the main mechanisms of formation of CTDs and FAEs.

In summary, our model (and the active model C \cite{Maryshev2019Dry,Maryshev2020}) differs significantly from the standard theory of active nematics \cite{ngo_large-scale_2014}, since it contains density-dependent corrections and higher order terms. Without such modifications the standard active nematic model is unable to reproduce CTDs.

\subsection*{From CTDs to FAEs and bands} \label{sec:from_ctd_to_fae}

\begin{figure}[!t]
\centering
\includegraphics[width=\linewidth]{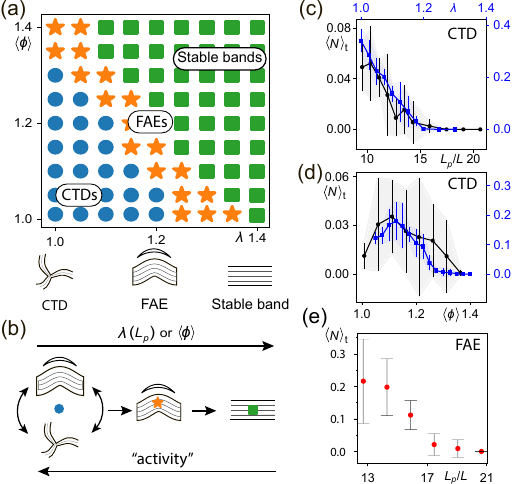}
\caption{Quantification of the occurrence of condensed defects and filamentous arc ejections in 
the agent-based simulations and the hydrodynamic model, plotted as a function of the indicated parameters. 
As the density and/or the persistence length (agent-based model) or $\lambda$ (hydrodynamic model) is increased, the most frequently observed type of collective state changes from CTDs to FAEs to straight lanes. 
(a) Phase diagram 
obtained from the numerical solution of the hydrodynamic model, which depicts the parameter regimes in which CTDs (blue dots), FAEs (orange stars), and straight nematic lanes (green squares) are the dominant structures. 
(For further details, see 
SI.)\
(b) Illustration of the three qualitatively different system behaviors as a function of the indicated parameters (upper arrow) or the strength of bending undulation (lower arrow). 
(c) Mean number of CTDs observed in the agent-based (hydrodynamic) model, plotted as a function of $L_p$ ($\lambda$).
(d) Mean number of CTDs observed in the agent-based (hydrodynamic) model as a function of $\langle \phi\rangle$.
(e) Mean number of FAEs observed in the agent-based model as a function of $L_p$.
}
\label{fig:fig3}
\end{figure}

Encouraged by the promising initial results shown by our hydrodynamic theory, we took advantage of the relative ease with which it can be used to determine the long-term behavior, and generated a ($\lambda, \, \langle\phi\rangle$) phase diagram [Fig.~\ref{fig:fig3}(a)]. 
As can be seen, at low values of $\lambda$ and $\langle \phi\rangle $, CTD formation dominates, while in areas of large $\lambda$ and $\langle \phi\rangle $ stable nematic lanes emerge. 
Between these regions lies a band of parameters where the system mainly exhibits FAEs. 
To test whether these findings obtained with the hydrodynamic model also hold for our agent-based simulations, we determined the average number of CTDs present at a given time in the agent-based simulation along one-dimensional lines of the ($L_p$, $\langle \phi\rangle $) phase space --- one along a  constant value of $\langle \phi\rangle $ and one along a constant value of $L_p$.
Reassuringly, the results for the agent-based simulations and hydrodynamic model are in good agreement [Figs.~\ref{fig:fig3}(c) and (d)]. 
We further checked the mean number of FAEs present in the agent-based simulations as a function of $L_p$ [Fig.~\ref{fig:fig3}(e)]; 
see SI for details. 
The observed decline in FAE frequency with increasing $L_p$ is consistent with the observations in the hydrodynamic model, where at high $\lambda$ no FAEs occur [cf. Fig.~\ref{fig:fig3}(a)].
Taken together, these results demonstrate that not only do the agent-based and hydrodynamic models share the same collective states, the frequency of these states also shows the same dependence on parameter changes.

The above relationships between model parameters and the occurrence of CTDs or FAEs can be related to the overall dynamic behavior (in short, ``activity'') of the system.
For both hydrodynamic and agent-based approaches, three distinct, qualitatively different dynamic states can be distinguished [Fig.~\ref{fig:fig3}(b)]. 

The \textit{first} of these is associated with very strong bending undulations of nematic lanes.
It occurs at low values of $L_p$/$\lambda$ or~$\langle \phi\rangle$ and is characterized by constant rearrangement of lanes [Movies S2,\,S3,\,S7,\,SI, Figs.~\ref{fig:fig2}(a),\,(b),\,(d) and (e)]: 
Lanes frequently collide leading to the formation of CTDs. In addition, system-spanning configurations of straight (or only slightly curved) lanes [cf. Figs.~\ref{fig:fig2}(c) and (f)], which may form randomly, are disrupted by undulations within a fairly short time. 
This is consistent with the observation that CTDs are the predominant phenomenon at low values of $L_p$/$\lambda$ and $\langle \phi\rangle$, respectively [Figs.~\ref{fig:fig3}(c), (d)].
Notably, FAEs can also be formed in this parameter regime following the emergence of short-lived system-spanning nematic lanes.
 
The \textit{second} dynamic state can be found at intermediate values of $L_p$/$\lambda$ or $\langle \phi\rangle$. 
In this regime, bending undulations are fewer and less pronounced, resulting in straight (or only slightly curved) and system-wide lanes that are stable over long periods of time:
Elongated openings often appear in the lateral areas of the lanes, which develop into filamentous arcs
[Movies S4,\,S8,\,SI, and Figs.~\ref{fig:fig2}(c),(f) and middle panel of Fig.~\ref{fig:fig3}(b)]. 
This is in accordance with the observation that FAEs are the predominant phenomenon observed at intermediate values of $L_p$/$\lambda$ or $\langle \phi\rangle$ [Figs.~\ref{fig:fig3}(a) and (c)-(e)]. 
 
The \textit{third} dynamic state is associated with vanishing bending undulations at high values of $L_p$/$\lambda$ or $\langle \phi\rangle$. Here, straight and system-spanning configurations are stable and no openings develop in their lateral regions [Movies S5,\,S9,\,SI and right panel of Fig.~\ref{fig:fig3}(b)]. Consequently, neither FAEs nor CTDs are observed [Figs.~\ref{fig:fig3}(a) and (c)-(e)].

The tendency just discussed for the bending undulations to become weaker as the values of $L_p$/$\lambda$ or $\langle \phi\rangle$ are increased from low to high values can be rationalized by the following heuristic arguments.
With increasing $L_p$/$\lambda$ the Frank constant \cite{sato1996frank} grows, and the effective elasticity (or collective stiffness of the polymers) yields stronger penalties for orientational distortions. 
As a result, the bending instability weakens, as described above.
The hydrodynamic model has allowed us to verify this hypothesis: upon varying the elastic constant $\kappa$ (independently from other parameters), we observe that weak elasticity favors the formation of CTDs, while a strong one yields stable bands. 
As the density $\langle \phi\rangle$ is increased (for a given and constant system size), a further effect contributing to higher stability of lanes is that a system-spanning nematic band occupies a growing fraction of space, i.e., the bands become wider while the bulk density remains largely the same [cf.~SI].
Since broader bands are less susceptible to a bending instability, an increase of $\langle \phi\rangle$, as discussed above, leads to the decay of defect formation.

An interesting aside can be mentioned here in the context of varying values of $\langle \phi\rangle$: for very small densities, close to the onset of order, both models show a drop in the observed CTD number [Fig.~\ref{fig:fig3}(d)], which is likely due to the fact that there is less mass within the ordered phase, and therefore not enough mass to form multiple curved bands necessary for lanes to collide and CTDs to be created.

Overall, the formation of condensed defects and filamentous arc ejections are both strongly linked to the stability of the nematic lanes, i.e., to their propensity to exhibit a bending instability~ \cite{ginelli_large-scale_2010, grosmann_mesoscale_2016, Maryshev2019Dry, Maryshev2020,Cai2019}, which, in turn, can be externally controlled by tuning either $L_p$/$\lambda$ or $\langle \phi\rangle$.

\subsection*{Detailed structure of CTDs and FAEs} 

To better understand the structure of the CTDs forming in agent-based simulations, we studied the polymer flows through them in detail. 
To this end, we tracked the motion of each polymer as it passed through a condensed defect. 
This enables us to distinguish the polymer flows from one to another arm of a defect and investigate whether there is a relationship between the lateral position of individual polymers and their eventual direction of turning. 

\begin{figure}[!t]
\centering
\includegraphics[width=           0.97\linewidth]{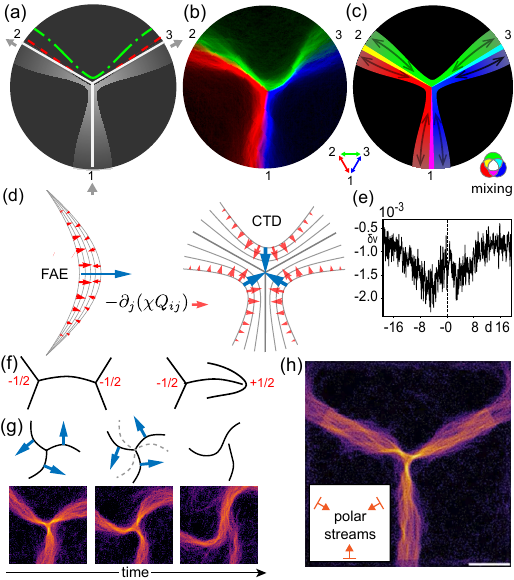}
\caption{
Structure of polymer fluxes through a CTD in agent-based simulations. (a) Schematic depiction of the flux density coming from one specific arm (the source arm, 1) into the target arms (2 and 3) in grey scale.
Solid white lines are the center lines of the arms.
The green dash-dotted line indicates the boundary of the total flux in the arms. The red dashed line is the boundary of the flux into the target arms. The two currents into a source arm mix only in a small region near the center line.
(b) Simulation data visualizing the information shown in (a). All polymers contributed by one specific arm and going into another specific arm or vice versa are depicted in the same color; i.e. all polymers derived from or entering arm 1 and entering or originating from arm 2 are colored in red.
The small mixing region of the different fluxes can be identified by the additive color mixing occurring when streams overlap (e.g. overlapping red and green fluxes lead to a yellow coloring of the flux-mixture). (c) Pictorial representation of the colored simulation data.
(d) Illustration of the anisotropic active flux in the hydrodynamic model. The flux leads to propagation in the indicated direction of the curved (bent) structures (left panel) and concentrates the density in a defect right panel.
(e) Average relative velocity change $\delta v$ 
of polymers as they pass through a CTD depicted as a function of the distance to the defect core $d$.   
(f) Schematic depictions of the typical mutual orientation of adjacent defects in phase-separated (left) and non-phase separated (right) active nematics. (g) Rotation and disintegration of a CTD. Blue arrows in schematic top panels show the direction of rotation. Dissolvement of defects is triggered by the detachment of one defect arm. (h) Artificially created CTD. By a proper placement of particle sources releasing polymers in a certain direction (orange arrows in the inset), an arbitrarily long existing CTD can be created at a definable position. [Parameters are (b) $\langle \rho \rangle L^2{=}3.5$, $L_p{=}11.1$, (e) $\langle \rho \rangle L^2{=}2.7$, $L_p{=}14.3$; see 
Appendix for further parameters.]}
\label{fig:fig4}
\end{figure}

Fig.~\ref{fig:fig4}(a) illustrates the flux from one arm of a defect (arm 1) into the two other arms (arms 2 and 3) [see Movie S6 SI for a representative flux recorded in an agent-based simulation].
The flux in each defect arm gets strongly compressed laterally in the vicinity of a defect core and then splits almost exactly at the centerline of the lane, while undergoing a sharp change in direction [Fig.~\ref{fig:fig4}(a)].
Symmetrically the same flux enters the defect from arms 2 and 3, resulting in the nematic flow structure depicted in Fig.~\ref{fig:fig4}(a) and (c). 
This also shows that the flows begin to mix again only at a greater distance from the center of the defect [cf. color mixing in Fig.~\ref{fig:fig4}(b) and (c)]. Hence, the overall topology often present at the birth of the defect [Fig.~\ref{fig:fig2}(b) and (e)] is preserved in the flow structure of the fully formed CTD as three barely intermingling nematic flows.

In addition, we investigated whether the velocity of the polymers is affected as they move through a CTD. As can be seen from Fig.~\ref{fig:fig4}(e), their speed remains almost unchanged and only a slowdown in the per mil range is observed. One can see two insignificant velocity drops corresponding to regions with the maximal density of polymers. Interestingly, in the immediate vicinity of the core of the defect, the particle velocity briefly returns to the average value, corresponding to particles inside the nematic band.  

We also studied the temporal evolution of FAEs and their occurrence over time. To this end, we periodically projected the density of a system in a configuration that allows the formation of FAEs onto one-dimensional slides and stacked them to obtain kymographs (see SFig.~5 SI). 
These reveal that the detachment of arcs accelerate over time. 
Further, they show that in the hydrodynamic model, due to no noise being present, FAE events occur at regular intervals, whereas in the agent-based simulations they form stochastically.

Having established the existence of CTDs and FAEs, and characterized them in our agent-based \textit{in-silico} experimental system, and having successfully introduced a hydrodynamic theory that faithfully reproduces the results of the simulations as well as providing access to the phase space of the observed pattern, we asked: why are these phenomena observed? What are the underlying mechanisms responsible for their formation?

To answer these questions, we leveraged the ability of the hydrodynamic model to provide access to single terms of its defining equations [Eqs.~(\ref{eqrho},\ref{eqQ})]. 
This analysis reveals that both the formation of dense defects and the movement of arcs have the same root cause, namely the anisotropic (``curvature-induced'') density flux \cite{ramaswamy2003active,simha2002hydrodynamic,narayan_long-lived_2007,Marchetti2013}, described by $-\partial_j(\chi Q_{ij})$ in Eq.~(\ref{eqrho}) in the hydrodynamic model.
This can be understood by plotting $-\partial_j(\chi Q_{ij})$ in the region of an FAE or a CTD; see the left and right panels of Fig.~\ref{fig:fig4}(d), respectively.
As can be seen, on opposite sides of the arcs the amplitudes of the fluxes are distinct. An effective ``active force'' acting on the concave side is greater than that on the opposite side, which leads to the movement of the bent band (or arc) in the corresponding direction [Fig.~\ref{fig:fig4}(d), left panel].

When three lanes meet, the same curvature-dependent fluxes concentrate polymers in the core of the resulting defect [Fig.~\ref{fig:fig4}(d), right panel]. This condensation is eventually balanced by the isotropic part of \eqref{eqrho} and particularly by steric repulsion of polymers.
To test this hypothesis, we set the excluded volume force $F$ (see SI)
to zero in our agent-based simulations. 
Observations in this case indicate that the formation of CTDs is reduced and that, when they form, they decay faster.
Thus, we conclude that formation of the dense defects is predominantly determined by the interplay between two counteracting processes: isotropic and anisotropic density fluxes.

In addition to the ``emergent'' way of obtaining CTDs just studied, in which spontaneously formed bands interact randomly and spontaneously condense into defects at stochastically distributed positions, we have sought a way to overcome this limitation by artificially generating and positioning CTDs.
In contrast to non-phase-separated systems --- where such an endeavor would involve the forced separation of a defect pair --- the way CTDs form spontaneously [Figs.~\ref{fig:fig2}(b),(e)] suggests that finding a way to position and form nematic lanes in suitable configurations could trigger the creation of a CTD.
In combination with the observation of polymer fluxes near a defect [Fig.~\ref{fig:fig4}(h)], we hypothesized that placing active polymer sources in a three-strand configuration should trigger the formation of three lanes that immediately condensate into CTDs. 
To test this prediction, we implemented the possibility to add such ``active particle throwers'' into our agent-based simulations and positioned them as described.
Indeed, we found that this way a CTD can be formed at a predetermined location where it persists for an arbitrary amount of time, cf. Fig.~\ref{fig:fig4}(h) and movie S10 SI.
This may be of potential application in cases where topological defects and/or high-density regions (in a low density background) need to be created and controlled with high accuracy.

\section*{Discussion}

In summary, we have used a combination of agent-based simulations and hydrodynamic theory to study pattern formation in phase-separated nematic active matter.
Our analysis shows that topological defects and nematic lanes, previously considered as two distinct and separate collective states, coexist and are tightly coupled.

We investigated the structure, formation and decomposition of CTDs in phase-separated systems. 
We observed that CTDs appear as characteristic collective excitations in a novel nonequilibrium steady state.
Moreover, the formation process of CTDs constitutes a new hierarchical condensation phenomenon.
Given the previously demonstrated and close connection of our agent-based algorithm to the actin motility-assay, a paradigmatic experimental model system, it is plausible to expect that CTDs will be observed in experimental active matter systems.
Below we discuss these observations step by step.

First of all, we characterized topologically charged structures, such as CTDs and  FAEs, for the first time observed in a phase-separated nematic system with self-propulsion.
It is apparent that CTDs  differ markedly from defects observed in homogeneous active matter, particularly in the dynamics of their formation and decay and in their spatial structure as well.

To begin with, CTDs upconcentrate density nearby their cores and condensate nematic fluxes. 
This condensation phenomena is interesting by itself, since the majority of experimental active matter systems show a depletion of particles in $-1/2$ disclinations, e.g., bacteria embedded in liquid crystals ~\cite{genkin2017topological} and cultures of neural progenitors~\cite{kawaguchi_topological_2017-1}.  
Weak density accumulation around the defects has been discussed in slightly inhomogeneous nematic \cite{mishra2014aspects}; 
however, in such systems, the $-1/2$ defects occur only during the transient and eventually disappear via annihilation with their $+1/2$ counterparts.

Similar CTDs, among other structures, were observed in parameter sweeps of the phenomenological toy model for mixtures of non-self-propelled microtubules and kinesin motors~\cite{Maryshev2020}. 
However, they were either transient or formed only under very special conditions (elasticity almost zero). 
In the latter case, the shape and the mechanism of formation of the defects were clearly different from the CTDs observed here.

In our case CTDs are typically formed by the collision of three curved nematic lanes that condense into a high-density three-armed structure, trapping the previously spatially distributed negative charge [Figs.~\ref{fig:fig2}(a),(d)].

One might think of comparing condensation to CTDs with the process of motility-induced phase separation (MIPS)~\cite{cates2015motility}.
However, the fundamental difference between the two is that CTDs are not associated with particle slowdown or prolonged residence of agents in high-density regions.
In addition, the formation of condensed defects provides a condensation mechanism for anisotropically shaped particles, which is not possible with MIPS~\cite{van2019interrupted}.
We may also argue that in MIPS the agents themselves condense into high-density clusters, while we observe the condensation of dynamical collective states (nematic lanes) into topological defects.

The mutual orientation of defects is also non-typical: we observe that two CTDs can be connected by a single nematic streamline (a filamentous bundle of polymers) [Figs.~\ref{fig:fig2}(a), \ref{fig:fig4}(f)], whereas in non-phase-separated active matter negative half-integer disclinations usually point towards a corresponding defect with the opposite charge $+1/2$ [Fig.~\ref{fig:fig4}(g)]~\cite{shankar2019hydrodynamics}.

The dynamic processes of defect decay in phase-separated and homogeneous active nematics are also clearly distinct. 
In homogeneous systems, pairs of defects with opposite charges annihilate each other~\cite{giomi_defect_2013, cortese2018pair}. In contrast, we find that CTDs do not annihilate with other defects, but disintegrate due to the undulating dynamics of the lanes that connect to the defect arms (Fig.~\ref{fig:fig4}(g) and Movie S3 SI). 
This means that the destruction of a negatively charged defect does not depend on the mobility or dynamics of a positively charged pair, rendering this process potentially easier to control.
In cases where all three lanes that connect to the respective arms have the same bending orientation (curvature of all either clockwise or anti-clockwise with respect to center), this decay takes place via an interesting process in which defects rotate before they dissolve [Fig.~\ref{fig:fig4}(g)].

Thus, CTDs not only emerge from ``collisions'' of nematic lanes, but also are connected by, and disassemble into them. 
Taken together, this leads to one of the main conclusions of our work, namely that the presence of CTDs constitutes
a novel nonequilibrium steady state which corresponds to a {\it dynamic equilibrium} between dense nematic lanes and condensed topological defects coexisting in a diluted background of disordered filaments.
This is reminiscent of other recent findings in active matter, in which a dynamical coexistence between patterns of different symmetry (nematic and polar) was observed~\cite{Huber2018,grosmann_particle-field_2020,denk_pattern-induced_2020-1}. During the persistent formation and subsequent decay of CTDs, those defects act as temporal capacitors of negative topological charge (i.e., the curvature on the boundaries of lanes gets temporarily trapped in a very small region of space) which eventually gets released again. 
It is well worth reiterating that this is a continuous cyclic phenomenon, not a transient one (unlike the defect formation observed in Ref.~\cite{mishra2014aspects}).

The most important factors that allow this nonequilibrium steady state to occur are probably the following.
First, since CTDs emerge from interaction of curved nematic lanes, a lateral undulation instability of nematic lanes --- as exhibited by our agent-based model --- is a basic prerequisite for their formation.
Another factor that is likely to favor the formation of CTDs is the nature of the interaction between the polymers (agents), which exhibit only weak mutual alignment and weak steric exclusion.
The latter, in particular, is likely to be a critical factor necessary for the high compression of polymer density during CTD formation.

Starting from a rigorously derived hydrodynamic model for self-propelled particles, we have generalized it to include higher-order phenomenological corrections. 
The resulting equations are reminiscent of a conceptual active model C~\cite{Maryshev2020}, but they include all terms arising from particle self-propulsion, which is an important additional feature here.
In particular, the hydrodynamic model presented here has many fewer degrees of freedom than the toy model presented in Ref.~\cite{Maryshev2020}, since the coefficients in front of all ``standard'' terms have a fixed relation among them.

This hydrodynamic theory provides additional insight into the physics of CTDs.
For example, it shows that density gradients play a crucial role through their coupling with the orientation field.
In particular, we consider density-dependent corrections of these coupling terms (controlled by the parameters $\chi_{\phi}$ and $\omega^a$), which typically disappear due to the linearization of terms around the mean value of density in the majority of hydrodynamic theories. 
We want to stress again that these additional terms, which are missing in standard theories of active nematics, are crucial for a proper description of the system, because without them CTDs are no longer observed.
We argue that strong phase separation (and the resulting large density gradients) inevitably amplifies the effect of higher-order coupling terms between the density and the orientation field on the dynamics.
For example, the bilinear anchoring  $\omega^a(\partial_i\phi)(\partial_j\phi)$ causes the nematic lines to closely follow the contour of the density field constituting a defect (SFig.~7 SI) and therefore can stabilize defects.
This is in line with the observation that a decrease in $\omega^a$ leads to a decrease in the number of defects (similar conclusion can be referred from \cite{Maryshev2020}). However, in our model CTDs still can be formed even if $\omega^a=0,\,\chi_{\phi}\neq0,\kappa_{\phi}\neq0$.

We firmly believe that the phenomena we found can also be observed in experiments, even though our study is purely theoretical.
The weakly aligning, self-propelled polymers simulation approach we base our study on, has previously shown not only excellent agreement with experiments, but was also able to predict then novel states that were later found in experiments~\cite{Huber2018}; thus it can be viewed, as elaborated in the introduction, as a computational version of an experimental system.
In light of this, we expect that the most promising experimental model system  that could allow observation of the new topological defects we predict is most likely the actomyosin motility assay~\cite{schaller_polar_2010, butt_myosin_2010, hussain_spatiotemporal_2013, suzuki_emergence_2017}.
This paradigmatic system not only satisfies the requirement of weakly interacting agents~\cite{suzuki_polar_2015,Huber2018}, but also offers the advantage of high particle numbers.
Previously, not only polar  waves~\cite{schaller_polar_2010} but also nematic lanes~\cite{Huber2018} have been observed.
This has been achieved by adding depletion agents that enable one to tune the strength as well as the symmetry of the interaction between the actin filaments. 
It is conceivable that similar and other changes in the design of the actin motility assay could be used to produce a weak and purely nematic interaction as used in our agent-based simulations.
For example, other depletion agents could be used and/or the properties of the surface to which the driving molecular motors are attached could be changed. 
Recently, the latter was indeed shown to have a direct impact on polymer interactions~\cite{sciortino_pattern_2021}.
Alternatively, CTDs could potentially be observed in other types of motility assays \cite{sumino_large-scale_2012-1,memarian_active_2021}.
Another intriguing possibility for observing the predicted CTDs is to directly produce a configuration of nematic lanes favoring the formation of CTDs by suitably structuring the surface used in the motility assay~\cite{turiv_polar_2020,sciortino_defects_2022}.

The deep understanding we gained about the formation of CTDs owing to the combination of agent-based simulation and hydrodynamic approach allowed us to find a way to generate them artificially (Fig.~\ref{fig:fig4}(h) and movie S10 SI). Given the availability of directed particle sources in an experimental system, the position of defects (and therefore the location of a domain of extremely high density) could be controlled with pin-point accuracy. 
This provides a new tool for cases where -1/2 defects and/or small regions of high particle density (in an overall dilute system) are needed at specific positions, e.g., to trigger specific processes such as cell death~\cite{saw_topological_2017-1} at definable points.  
Given the strong and controlled nature of the focusing of the fluxes in nematic lanes, this method could be termed ``active matter optics''.

Another important insight from the broader perspective of the active matter field is that
phase-separated active matter exhibits a hierarchy of emergent collective states.
Interaction between dense nematic lanes, considered as ``first-order'' collective states in active nematics, can lead to the formation of ``second-order'' collective states, here half-integer topological defects with an even higher density. 
A phenomenon which one can call ``hierarchical, alignment-induced phase-separation''.
It is reasonable to assume that similar effects may lead to new phenomena in other active systems with different symmetry, e.g., polar symmetry with polar waves as first-order collective states~\cite{chate_dry_2020, huber_microphase_2021}. 
Another class of systems in which higher-order collective states might emerge are active systems that are subject to external gradients~\cite{popescu_chemotaxis_2018}  
or signalling interactions between the agents~\cite{lavergne_group_2019,alex_preprint_2022}.

A promising extension of our present investigations are active foams. 
In this state of active matter, which has recently received increasing attention~\cite{Maryshev2020, nagai_collective_2015-1, ventejou2021susceptibility, lemma2022active}, dense ordered bands assemble into actively reforming cellular networks. 
Indeed, in preliminary simulations of the hydrodynamic theory, we have identified parameter regimes in our model where we observe active foams: CTDs are more frequent, interconnected, and persist for longer times.
Thus, the formation of active foams in active nematics seems very plausible, but a thorough investigation of the entire phase space in the agent-based model is computationally demanding and will be reserved for a future study.

\section*{Author Contributions}
T.K., I.M., and E.F. designed the research, performed research, analyzed data, and wrote the paper.

\section*{Conflicts of interest}
There are no conflicts to declare.

\section*{Acknowledgements}
We acknowledge financial support by the Deutsche Forschungsgemeinschaft (DFG, German Research Foundation) through the Excellence Cluster ORIGINS under 
Germany's 
Excellence Strategy (EXC-2094-390783311) and through Project-ID 201269156 -
Collaborative Research Center (SFB) 1032 - Project B2.
IM acknowledges European Union's Framework Programme for Research and Innovation Horizon 2020 (2014-2020) under the Marie Sk\l{}odowska-Curie Grant Agreement No. 754388 (LMU Research Fellows) and from LMUexcellent, funded by the Federal Ministry of Education and Research (BMBF) and the Free State of Bavaria under the Excellence Strategy of the German Federal Government and the L\"ander.

\section*{Appendix}

\subsection*{Agent-based simulation method}
We now describe our agent-based simulation model. 
Please also refer to the SI and the Supplemental Materials of Refs.~\cite{Huber2018,huber_microphase_2021} for more details. 

In our systems we simulate $M$ polymers, each of length $L$.   
Orientational diffusion causes the tip of each polymer to perform a persistent random walk. Upon collision with another polymer, local interaction causes the tip to gradually align with its direction. 
Attached to the polymer tips are tails that just follows the path that is outlined by the tip.  

This dynamics mimics the behavior of actin filaments in actomyosin motility assays \cite{Huber2018, huber_microphase_2021}, in which polymers move in a snake-like fashion over a lawn of motor proteins and motion orthogonal to the contour is suppressed~\cite{schaller_polar_2010, suzuki_polar_2015}. 
Here we use purely nematic interactions between polymers which are primarily tuned by the nematic alignment amplitude $\alpha_n$ that allows for a continuous variation of the rate of alignment.  

\subsection*{Parameters}
If not stated otherwise, we used the following model parameters: discretization $N\,{=}\,5$, polymer aspect ratio $L/d\,{=}\,21$, nematic alignment strength  $\alpha_n\,{=}\,0.126\approx7.2^\circ$ and a periodic simulation box of length $L_\text{box}\,{=}\,162.5L$. 
The velocity $v^{(n)}$ of each polymer is randomly drawn from the interval $[0.75,1.]v_0$.
We started simulations with random initial conditions, i.e.\ randomly oriented polymers were placed at random positions in the simulation box.
Time is measured in units of $L/v_{0}$, where $v_{0}$ is the maximal velocity of a free polymer. 
Density in Figs.~\ref{fig:fig2}(a)-(c) and \ref{fig:fig4}(g)-(h) is time-averaged for better visibility, with averaging times of 159 for Fig.~\ref{fig:fig2}(a) and 16 for Figs.~\ref{fig:fig2}(b)-(c) and \ref{fig:fig4}(g)-(h). 
Note that the system shown in Fig.~\ref{fig:fig4}(h) does not have the usual periodic boundary conditions. Rather, the particles crossing the boundaries are moved either to a random position along a boundary with random orientation or to one of the particle sources. The ratio of these two possibilities is chosen so that the particle flux from the sources is kept constant.

\subsection*{Continuous theory}

We numerically investigate Eqs.~(\ref{eqrho},\ref{eqQ}) under periodic boundary conditions by using finite differences of second order \cite{AbramowitzStegun} on a $300\times300$ grid with the spatial resolution $\delta x = 0.5$. 
The time integration was performed via a second-order predictor-corrector scheme  with time step $dt = 10^{-2}$.
We use the parameter values $\beta=0.05,\,\kappa_{\phi}=0.2,\,\omega^a=-0.5, \,\chi_{\phi}=0.4,\,\nu_{\phi}=1$.
Unless explicitly stated, we initialize simulations from an isotropic uniform state 
with a small amount of noise. To make time and space dimensionless we rescale them by setting the rotational diffusion coefficient and $\mu_{\rho}$ equal to unity.

\onecolumn
\newpage


\captionsetup*{format=largeformat}

\section*{Supplementary Information}
\section{WASP simulation method} \label{sec:SI_wasp}

In this section we provide a brief summary of the agent-based simulations. 
The focus will be on the aspects most relevant for the current study. 
For a detailed description of the WASP simulation setup, please refer to the supplemental materials of Refs.~\cite{Huber2018, huber_microphase_2021}.   

In the agent-based simulations, we consider $M$ polymers moving on a flat substrate (in two spatial dimensions).
Each polymer $n$ consist of $N$ spherical joints $j$ which are located at a positions $\mathbf{r}_j^{(n)}$ (with $j \,{\in}\, \{ 0, 1, \ldots, N \,{-}\, 1 \}$, where the polymer tip is denoted by $j\,{=}\,0$).  
The direction of a polymer's tip is denoted by $\mathbf{u}_0^{(n)}$ and its motion is described by:
\begin{align} 
\label{seq:wasp_basic}
	\partial_t \mathbf{r}_0^{(n)}
	&= v^{(n)} \, \mathbf{u}_0^{(n)} -\mathbf{F_\text{rep}}
	 = v^{(n)}
	 \left(
	 \begin{array}{c}
	 \cos \theta_0^{(n)} \\ \sin \theta_0^{(n)} 
	 \end{array}
	 \right)-\mathbf{F_\text{rep}} 
	 \, . 
\end{align}
Here $\mathbf{F_{\text{rep}} }$ describes a weak repulsion force (see \eqref{seq:repFrc}) acting on a polymer head while in contact with the contour of another polymer. 
$\theta_0^{(n)}$ denotes the orientation of a polymer and $v^{(n)}$ its free speed. 
For this study, the speed of each polymer was chosen at random from a continuous uniform distribution in the interval $[0.75, 1] v_0$, where $v_0$ denotes the maximal velocity of a free polymer (see section S\ref{si:sec_fil_Vel} for further details on this velocity dispersion).

The orientation of a polymer's head evolves in time according to
\begin{align} \label{seq:wasp_angle}
	\partial_t \theta_0^{(n)}
	&= 
	- \frac{\delta \tilde{H}_0^{(n)}}{\delta \theta_0^{(n)}}
	  + \sqrt{\frac{2v^{(n)} }{L_p}} \, \xi \, ,
\end{align}
where $\xi$ is random white noise with zero mean and unit variance with the magnitude of the noise given by the prefactor.
This implies that individual polymers perform a persistent random walk with a path persistence length of $L_p$. 
$\tilde{H}_0^{(n)}$ sets the---in this study purely nematic---torque caused by interactions with other polymers.

Before we come to a description of $\tilde{H}_0^{(n)}$, it will proof useful to introduce several other quantities. 
The first is the distance vector 
\begin{equation}
	\Delta \mathbf{r}_{nm}
	= 
	\bigl(
	\mathbf{r}_0^{(n)} - 
	\mathbf{r}^{(m)}
	\bigr)_\text{shDist}
	\, . 
\end{equation}
This vector connects the tip of a polymer $n$ with the position of an adjacent polymer's (denoted by $m$) contour that has the shortest possible distance. 
The local orientation of the contour of the adjacent polymer $m$ is given by $\theta_j^{(m)}$, which corresponds to the orientation of the polymer segment $j$ of polymer $n$ to which $\Delta \mathbf{r}_{nm}$ connects. 
Second, if a polymer is interacting with several polymers at a time, we define a weighted average direction of the connecting vectors:
\begin{align} \label{seq:q1}
	\Delta\widetilde{\mathbf{e}}_{n}
	:=  
	\sum_{m} 
	C\left(
	\left|\Delta \mathbf{r}_{nm}\right|
	\right) 
	\frac{\Delta \mathbf{r}_{nm}}
		 {\left|\Delta \mathbf{r}_{nm}\right|}
	\, .
\end{align} 
Here $C\left(
	\left|\Delta \mathbf{r}_{nm}\right|
	\right)$ is a weighting factor accounting for the assumption that a more distant polymer contributes less to an interaction. 
It is given by 
\begin{align} \label{seq:ramp}
	C\left(
	\left|\Delta \mathbf{r}_{nm}\right|
	\right) 
	=
	\left
	\lbrace
	\begin{array}{cc} 
		0 & \text{if}\ 	
		\left|\Delta \mathbf{r}_{nm}\right|{>}d \\ 
		(d-	\left|\Delta \mathbf{r}_{nm}\right|)/d & \text{else} 
	\end{array} \right. 
	\,, 
\end{align}
where $d$ defines the interaction radius. 
Using the orientation of the averaged connecting vector $\tilde{\theta}_{n}$, we define an averaged nematic impact angle as $\Delta \tilde{\theta}^{(n)}_{n} \,{=}\,  \theta_0^{(n)} \,{-}\, \tilde{\theta}_{n}$. 
Equipped with these definitions we are now in a position to write down the alignment potential as 
\begin{align} \label{seq:wasp_align}
    \tilde{H}_0^{(n)}
    := 
    \frac{\alpha_n v_0}{d} \cos (2\Delta \tilde{\theta}^{(n)}_{n}) |\Delta\widetilde{\mathbf{e}}_{n}|
    \, ,
\end{align}
where the overall amplitude of the alignment is set by the absolute value of the weighted connecting vector, combined with the nematic alignment strength $\alpha_n$.

The repulsion force $\mathbf{F_\text{rep}}$ in \eqref{seq:wasp_basic} is given by
\begin{align} 
\label{seq:repFrc}
\mathbf{F_\text{rep}}
	&= 
	-s \sum_{m} 
	C
	\left(
	\left|\Delta \mathbf{r}_{nm}\right|
	\right)	
	 \frac{\Delta \mathbf{r}_{nm}}{\left|\Delta \mathbf{r}_{nm}\right|}
	\, ,
\end{align}
which is used to prevent unphysical aggregation of polymers. It is assumed to be weak with $s\,{=}\,0.05$. 
Filaments in actomyosin motility assays are observed to conduct a trailing motion, where the tail of a polymer follows the movement of the tip~\cite{schaller_polar_2010, butt_myosin_2010, hussain_spatiotemporal_2013, suzuki_polar_2015,Huber2018}. 
To emulate this behaviour, tail joints move according to 
\begin{align} \label{seq:tail_force}
	\partial_t \mathbf{r}_j^{(n)}
	=
	 K_s \, 
	\left(
	\left| \mathbf{r}_{j}^{(n)}{-}\mathbf{r}_{j-1}^{(n)}\right| - b
	\right) \,
	\frac12
	\left(
	\mathbf{u}_{j+1}^{(n)}+\mathbf{u}_j^{(n)}
	\right)
	\, .
\end{align}
Here, the second part of the equation, $\frac12\, 
	(
	\mathbf{u}_{j+1}^{(n)} \,{+}\, \mathbf{u}_j^{(n)}
	)$, ensures the movement to be in the direction of the average of the segment's orientations that are adjacent to joint $j$.
The remainder of \eqref{seq:tail_force} corresponds to a linear (Hookian) restoring force with spring coefficient $K_s \,{=}\, 200$ that ensures an average length $b$ of the cylindrical segments between bonds.

\section{Onset of nematic patterns}
In this section we provide further information on how the phase diagram shown in Fig.~1(c) of the main text was obtained.

To determine the density $\rho_n$ as a function of $L_p$ above which nematic patterns are formed, we performed exploratory simulations in the phase space spanned by the (reduced) global polymer density $\langle \rho \rangle L^2$ 
and the persistence length $L_p$.
To guarantee that the dynamics has reached a steady state we ran these simulations for a time $15 \, 873$ which is much larger than the initial timescale $t_0 \approx 100$ it takes for a system to reach the quasi-stationary, disordered state \cite{huber_microphase_2021}. 
Figure~\ref{sfig:onset_nem} shows the results of the \textit{in silico} parameter scans in density at a set of fixed values for $L_p$: The blue triangles and red squares correspond to steady states where we visually observed nematic patterns or a disordered state, respectively. 
To determine the phase boundary $\rho_n (L_p)$ we fitted a function ${f_{\rho}(L_p) = a/L_p}$ (with $a$ as free fitting parameter) to the data points with the lowest density that still exhibited nematic order [solid line in Fig.~\ref{sfig:onset_nem}].

The shape of the boundary line is dictated by the interplay between two counteracting effects: density-dependent, interaction-induced ordering and rotational diffusion. 
The former increases linearly with density increase, and above the critical value of density, spontaneous ordering begins to predominate over diffusion.
Thus, the critical density is proportional to rotational diffusion coefficient and therefore $\propto L_p^{-1}$ in our case.
We take $f_{\rho}(L_p)$ as an approximation to the density corresponding to the onset of nematic patterns, $\rho_n (L_p)$. 
 
To further test whether this is a satisfactory approximation for the phase boundary, we ran ten independent simulations at a density corresponding to $\rho_n$ [cf. dots in Fig.~1~(c) of the main text] and further ten at $0.9 \, \rho_n$ for several different $L_p$ for a twice as large simulation time of $31 \, 746$.
All simulations at $\rho_n$ formed ordered patterns, while none at $0.9 \, \rho_n$ did, affirming that $f_{\rho}(L_p)$ adequately approximates the position of the isotropic-nematic transition.

\section{Defect detection} \label{si:cd_detect}

In this section, we explain the algorithms we used to identify topological defects in simulations of both the hydrodynamic theory and the agent-based model.

To algorithmically detect $-1/2$ defects in 
both approaches, we took advantage of the fact that inside a defect core the topological charge density $q$, defined as \cite{Blow2014}
\begin{align} 
\label{seq:q_def}
q = \frac{1}{4 \pi} \left( \partial_{x}\hat{Q}_{x a}\partial_{y} \hat{Q}_{y a} - \partial_{x}\hat{Q}_{y a}\partial_{y} \hat{Q}_{x a} \right),
\end{align} 
has a very large negative value (with $\hat{Q}{=}Q/\rho$ and $Q$ defined as in \eqref{eq:def_rho_p_q}), whereas in other regions of space its absolute value is much smaller (cf. lower right pane of Fig. 2(a) and (d) of the main text). We exploit this fact and define any contiguous region of space in which $q$ falls below a certain threshold value $q_{\text{thrs}}$ as one $-1/2$ defect.

The position of $ -1 / 2 $ defects in the \textit{agent-based model} is obtained in the following way. 
Please first note that the main purpose of the data from the agent-based simulations in Fig.~3(c)-(e) is to qualitatively confirm the trend observed in the hydrodynamic model. To quantify the data with a high degree of precision would require averaging over large ensembles, which would be numerically prohibitively demanding given the very long time scales on which the observed phenomena occur. 

The total runtime of each simulation was $142\, 857$ (which is much longer than the dynamics of undulations; cf. Movie S1 and S2), from which we cutted an initial transient (cf. section S\ref{si:fae_meas}) before starting the measurement. 
For each value of $L_p$/$\langle \phi \rangle$ we averaged over ten independent simulations.

To obtain $q$ in agent-based simulations, we rasterized space into a grid with a grid spacing of $\Delta x = 0.3$, which is small enough to resolve the structure of a defect (note that the qualitative agreement between the agent-based simulations and hydrodynamic model, shown in Fig. 3 of the main text, does not depend on the exact choice of this and the following numerical parameters). 
We used the orientations $\theta_0^{(n)}$ of polymer tips residing inside each grid point at a given time to calculate a local value of $\hat{Q}$ using \eqref{eq:def_rho_p_q}. To suppress noise due to stochastic particle fluctuations, we further averaged over a time span of $15.9$, which is much shorter than density rearrangements due to bending undulations. 
With this we obtained $q(\mathbf{r}, t)$ using \eqref{seq:q_def}. 
We chose $q_{\text{thrs}}\,{=}\,\,{-}\,0.032$, which is much lower than typical values of $q$ outside defects. 
Additionally, to avoid classifying small and short-lived density peaks that occur sporadically in the simulations as CTDs, we heuristically filtered them out by requiring the charge density to be below $q_{\text{thrs}}$ for a time of at least 159 for a CTD to be detected.

The \textit{hydrodynamic model} allows by construction a direct access to the $Q$-tensor, which allows a direct calculation of the function $ q $, given by Eq.~\ref{seq:q_def}. The positions of $ -1 / 2 $ defects are defined as local minima of the function $ q $ and, for consistency, the same value of $q_{\text{thrs}}$ is used as for the agent based simulations.
For the measurements in the hydrodynamic model, we discarded the data collected in the first half of the simulation runs in order to avoid any influence of initial transients. 
To generate the data shown in Fig.~3 (a), we classified all runs in which CTDs were detected to be CTD-dominated (blue dots in Fig.~3 (a)). Distinction between FAEs and stable bands was made via visual inspection.

\section{Flux measurement through defects}

In the main text, we studied the mass flow through a defect as well as the speed of particles during a CTD passage; see Figs.~4(b) and 4(e), respectively. 
To this end, we needed detailed information about the position and velocity of particles as they transitioned from one arm of a defect to another. 
To determine these quantities, we leveraged the possibility offered by the agent-based simulations to access the position of each individual polymer at any given point in time. 

In order to be able to deduce that a given polymer has transitioned from one arm of a defect to another one, several things have to be known. 

First, one has to find a criterion which allows to algorithmically determine if a polymer is pertinent to a given arm at a given time.
For this we used the following heuristics: 
Over each arm of a defect we placed a round ``classification area'', which is large enough to cover the full width of the nematic lane (blue regions in Fig. \ref{sfig:filter_regions}, diameter $22\,L$). 
The positions of the classification areas were chosen such that they roughly coincided with the area where the nematic lanes recovered their full width (midpoint distance of classification areas to defect: $26\,L$ in Fig. \ref{sfig:filter_regions}). 
Every polymer being inside one of these regions is classified as pertinent to the given defect arm.

Second, one has to find a criterion that allows to make a determination as to the origin of particles that have been classified as belonging to a particular arm. 
For this we introduced an additionally classification area which encompasses all parts of the simulation box being further away from the defect core than a specific distance, cf. orange region in Fig. \ref{sfig:filter_regions} (distance to defect: $40 \,L$).
(Note that the black colored area does not pertain to any classification area.) 
After this partitioning, we measured the currents from one region to another with the below described heuristics.
We did this for a time span sufficiently long enough that many particles can travel from one blue region to another blue region (cf. Fig. \ref{sfig:filter_regions}), but short enough such that bending undulations do not change the position of the individual lanes significantly.  
Data in Fig.~4(b) averaged over $159$, Fig.~4(e) averaged over $4 \, 019$ trajectories in a time of $317$.

For the flux measurement heuristics, we each assigned a unique identifier $id$ to every classification area.
We then checked in short intervals of 0.16 for every polymer $i$ if its position coincided with one of the classification areas.
If this was the case, polymer $i$ was assigned the identifier of the region and the time of assignment $t_{\text{assign}}$ was saved. 
If polymer $i$ already had a different identifier $id'$ assigned (and hence also a different $t_{\text{assign}}'$), this meant that it had traveled from another classification area into the current region (without crossing a third region in the meantime). 
In such a case, we stored the pairs of tuples $(id', t_{\text{assign}}')$ and $(id, t_{\text{assign}})$, which allow (combined with with the also saved information of the position and speed of every polymer at every interval) to reconstruct the path polymer $i$ has taken propagating from region $id'$ to $id$. Subsequently, we replaced the assigned identifier and assignment time of polymer $i$ with that of the current region and the current time and continued the simulation. \\

\section{Dispersion in the polymer velocity} \label{si:sec_fil_Vel}
Most studies of active matter assume the speed of agents to be constant and uniform \cite{chate_dry_2020}. Yet, experiments of the actin motility assay show actin filaments to have a broad distribution of velocities~\cite{Huber2018}. 
To take into account the effects of such a velocity dispersion, we drew the assigned speed of polymers from a distribution (cf. Section S\ref{sec:SI_wasp} of this Supplemental Material).
We have found that the introduction of such a velocity dispersion does not hinder the formation of nematic lanes. 
To additionally check whether particles that possess different free velocities behave differently on the level of macroscopic structures---for example by causing an effective sorting of particles into spatially separate populations, where only relatively fast/slow particles form part of patterns---we subdivided the system into a grid with a grid spacing of $\Delta x = 0.3$ and determined for each grid-cell the locally averaged $\langle v^{(n)}\rangle$ of particles inside a simulation exhibiting nematic lanes and CTDs. 
Any local accumulation of fast/slow particles would lead to a different value of $\langle v^{(n)}\rangle$ when compared to the global average $\langle v^{(n)} \rangle_{\text{glob}}$.
As can be inferred from Fig.~\ref{sfig:local_avg_vel}, the system is well mixed (up to random fluctuations) with respect to polymer velocities.
We further found that the introduction of a velocity dispersion prevented the decay of purely nematic patterns into oppositely propagating polar waves (cf. Ref~\cite{Huber2018}), which hence seems to be an artefact of the assumption of equal and uniform velocities.

\section{Width of nematic lanes}

As discussed in 
the main text, 
we measured the width of nematic lanes as a function of density $\langle \phi\rangle $ in both the agent-based simulations and the hydrodynamic model (at a constant system size). 
To this end, we performed several simulations at different polymer densities but at a fixed persistence length (resp. several realizations of the hydrodynamic model at different $\langle \phi\rangle $ and fixed $\lambda$). 
After these systems had reached a configuration in which they exhibited a single straight lane, we measured the width of the band and the average density $\langle \phi\rangle_\text{bg}$ in the disordered background. 
(The width is determined by averaging the density of the system along the axis of the straight lane, which results in a one dimensional density profile.
The width of the lanes in the hydrodynamic model is then defined as the distance between the two points with the maximal gradient of this curve, which can easily be obtained due to the absence of noise. 
In the agent based simulations the lane width is heuristically defined as the width of the region where this profile exceeds the threshold of three times $\langle \phi\rangle_\text{bg}$.)
As shown in Fig. \ref{sfig:width_nem_lane}, the thickness of the lanes grows linearly with density in both the agent-based simulations and hydrodynamic model, while the density of the disordered background remains constant.

\section{FAE detection} \label{si:fae_meas}
In this section we describe the procedure we used to measure the mean number of FAEs present at different parameter regimes in the agent-based simulation (Fig.~3(e) of the main text).

For this we logged the formation of every FAE in the investigated systems; the most reliable method for detecting FAEs turned out to be manual inspection of simulation videos. 
To obtain the mean number of FAEs present, we divided the total lifetime of all detected FAEs in the system by the total observation time. 
For every investigated $L_p$ in the agent-based simulations, we averaged over ten independent simulations, which each ran for a time of $142 \, 857$.

It is worth to note that agent-based simulations started in a parameter regime in which systems predominantly exhibit FAEs or stable lanes (i.e., high $L_p$; see also section ``From CTDs to FAEs and bands'' in the main text), 
do not immediately form straight lanes at the onset of pattern formation, but frequently at first dwell in a state of high activity (cf. left panel of Fig. 3(b) in the main text) in which no FAE can develop.
We measured the duration of this initial transient (``dwell-time'') and found that it is shorter than a time of $70\, 000$ in more than ninety percent of the cases.

We discarded this initial time span in the measurements of the mean numbers of CTDs (cf. section S\ref{si:cd_detect}) and FAEs present to rule out any influence of the initial transient on the results.

Further, we studied the temporal evolution of filamentous arc ejections. 
The motion of a separating arc in the agent based and the hydrodynamic model, can be visualized using a kymograph of the density projection shown in Fig.~\ref{sfig:kymograph_fae}. 
As can be inferred from the bending of the lateral extrusions, the separation process of the arcs starts slowly and continues to accelerate until complete ejection and eventual dissolvement of the arc.

\section{Hydrodynamic model}
\label{sec:hydrodynamic_model}

To provide the motivation of our hydrodynamic model we start form the general form of the evolution equation for the probability distribution function $P(\mathbf{r},\theta,t)$:
\begin{align}
    \partial_{t} P(\mathbf{r},\theta,t)
    =
    - L_p \partial_i \big[ n_i P(\mathbf{r},\theta,t) \big]
    + \partial_\theta^2 P(\mathbf{r},\theta,t) +\text{interactions}
    \, ,
\label{sec:pdf}
\end{align}
where ${\mathbf{n}=(\cos\theta,\sin\theta)}$ is director vector, and $L_p$ is the path persistence length of the polymers.
Time is measured in units of the diffusion coefficient.
Note that we only consider rotational diffusion and neglect translational diffusion. 
In the following the space and time dependencies of the probability density  are suppressed for brevity.
Contribution from the interaction between the polymers can be introduced in the form of collision intergrals in the Boltzmann ansatz \cite{bertin_boltzmann_2006,
bertin_2009,
Peshkov2012,
peshkov_boltzmann-ginzburg-landau_2014}, or by using the gradient of the interaction-induced current in a Smoluchowski approach \cite{ baskaran2010nonequilibrium}.

We define the particle density $\rho$, the polarity vector $\mathbf{p}$, and the nematic Q-tensor as the first three moments of the probability distribution function:
\begin{align} \label{eq:def_rho_p_q}
    \rho  
    :=
    \int_{0}^{2 \pi} \text d \theta \, 
    P (\theta) 
    \, ,\qquad
    p_i  
    := 
    \int_{0}^{2 \pi} \text d \theta \, 
    n_i P(\theta) 
    \, ,\qquad
    Q_{ij} 
    := 
    \int_{0}^{2 \pi}\text d \theta \, 
    \left(
    2n_{i} n_{j}-\delta_{i j}
    \right) P (\theta) 
    \, ,
\end{align}
where the subscripts $i$ and $j$ denote the Cartesian components and $\delta_{ij}$ represents the Kronecker delta. 
It is convenient to consider Fourier harmonics of the probability distribution function:
\begin{equation}
   P(\mathbf{r}, \theta)=\sum_{k=-\infty}^{\infty} P_{k}(\mathbf{r})  e^{i k \theta}.
 \end{equation}
According to their definitions, $\rho$ , $p_i$, and $Q_{ij}$ can be expressed via Fourier harmonics as follows:
\begin{subequations}
\begin{align}
    \rho  
    &=
    2 \pi P_{0} 
    \, , \\
    p_i  
    &=
    \pi
    \big(
    (P_{1} +P_{-1} ), \mathrm{i}(P_{1} -P_{-1} )
    \big)
    \, , \\
    Q_{ij} 
    &= 
    \pi
    \big(
    (P_{2} +P_{-2} ), \mathrm{i} (P_{2} -P_{-2} )
    \big)
    \, ,
\end{align}
\end{subequations}
where the symbol $\mathrm{i}$ denotes the imaginary unit.

By introducing the projection onto the $m^{\text{th}}$ harmonics of $P$:
\begin{equation}
    \overline{(\ldots)}^{\,m}
    :=
    \frac{1}{2 \pi} 
    \int_{0}^{2 \pi} \text d \theta \, e^{-\mathrm{i} m \theta}(\ldots) 
    \, ,
\end{equation}
one obtains the following contributions from the advective and diffusive parts of \eqref{sec:pdf} to the evolution equations of the $m_{\text{th}}$ Fourier harmonics ($P_{m}$):
\begin{align}
    \partial_{t} P_{m} 
    &=
    -m^{2} P_{m}
    -\overline{L_p\partial_i(n_iP(\mathbf{r},\theta))}^{\,m}
    \nonumber\\
    &=
    -m^{2} P_{m}
    - L_p\frac{1}{2}
    \bigg[
    \partial_x\sum_kP_k (\delta_{k,m-1}+\delta_{k,m+1})
    +\partial_y\sum_k P_k (\delta_{k,m-1}-\delta_{k,m+1})/\mathrm{i}
    \bigg]
    \, .
\end{align}
In terms of the collective variables this can be rewritten as:
\begin{subequations}
\begin{align}
    \partial_{t} \rho 
    &= 
    - L_p\partial_ip_i 
    \, , 
    \\
    \partial_{t} p_i 
    &=
    -p_i-\frac{ L_p}{2}\partial_i\rho +\frac{L_p}{2}\partial_jQ_{ij} 
    \, ,
    \\
    \partial_{t} Q_{ij} 
    &=
    -4Q_{ij}
    -\frac{L_p}{2}
    \big[
    \partial_ip_j+\partial_jp_i-\delta_{ij}\partial_kp_k
    \big]
    \, .
\end{align}
\end{subequations}
Note, that we imply summation for repeating indices following the Einstein convention. 
Since we consider a system with purely nematic interactions, the polar order decays on short time scales for all strengths of self-propulsion. 
Thus, the polarity field $\mathbf{p}$ equilibrates fast and can be eliminated adiabatically to arrive at dynamic equations for the density $\rho$ and Q-tensor alone. 
We find after rescaling time by a factor of 4:
\begin{subequations}
\label{advdiv}
\begin{align}
    \partial_{t} \rho
    &= 
    \lambda^2\Delta\rho
    + \lambda^2\partial_i\partial_jQ_{ij}
    \, ,\label{Eq:density}\\
    \partial_{t} Q_{ij} 
    &= 
    -Q_{ij}
    +\frac{\lambda^2}{2}\Delta Q_{ij}
    +\lambda^2
    \big[
    \partial_i\partial_j\rho
    \big]^\text{st}
    \,,\label{Eq:Q}
\end{align}
\end{subequations}
where we have introduced the parameter ${\lambda:=L_p/(2\sqrt2)}$,  $\Delta=\partial_i\partial_i$ denotes the Laplace operator, and $[...]^\text{st}$ indicates the symmetric and traceless part of the expression.

We now discuss the physical meaning of each term on the RHS of
Eqs.\,(\ref{advdiv}).
The first term in the density equation Eq.\,(\ref{Eq:density}) acts like effective translational diffusion, despite the fact that it is actually coming from the single particle advection (note, that the real translational diffusion is neglected in our model).   
The second term in equation Eq.\,(\ref{Eq:density}) represents anisotropic flux of material along the nematic order. This term enhances diffusion along the direction of the eigenvector of $Q_{ij}$
corresponding to its positive eigenvalue, and suppresses it along the perpendicular direction. It also can be treated as \textit {curvature-induced} flux, since it disappears in a uniformly ordered state. 

The first term in the evolution equation of the nematic tensor Eq.\,(\ref{Eq:Q}) is due to the thermal rotational diffusion. If there were no interaction between polymers, the action of this term would lead to disordering.
The second term in Eq.\,(\ref{Eq:Q}) penalizes the distortion of $Q_{ij}$ and represents the elasticity in terms of liquid crystal theory. 
The last term of Eq.\,(\ref{Eq:Q}) provides the coupling between the equations.  It can be treated simply as an anisotropic diffusive contribution. But it also introduces ``aligning torque'' by changing the orientation of nematic order in the presence of the density gradients.

Finally, besides the diffusion- and advection-related terms we need to add interaction-induced contributions. 
Inspired by Refs.\,\cite{Maryshev2019Dry,Maryshev2020} we also introduce the following terms to describe the nematic interactions of the polymers:
\begin{subequations}
\label{colonly}
\begin{align}
    \partial_{t} \rho 
    &=
    \cdots
    +
    \tilde\nu_{\rho} \Delta\rho^2 
    +
    \tilde\chi_{\rho} \partial_i\partial_j(\rho Q_{ij}) 
    \, , \label{rho_corr}\\
    \partial_{t} Q_{ij} 
    &=
    \cdots
    + 
    \tilde\alpha \rho Q_{ij}
    -
    \tilde\beta   Q^2Q_{ij}
    +
    \tilde\kappa_{\rho} \langle\rho\rangle \Delta Q_{ij}
    +\tilde\omega^a
    \big[
    2\partial_i\rho\partial_j\rho
    \big]^\text{st}
    \label{Q_corr}
    \, .
\end{align}
\end{subequations}
The $\tilde\nu_{\rho}$-related term in Eq.\,(\ref{rho_corr}) comes from the excluded volume interactions between the polymers (however an analogous term occurs due to the ``collision" of polymers, e.g., see Ref.\,\cite{Maryshev2019Dry}). 
The last term in Eq.\,(\ref{rho_corr}) is an interaction-induced flux representing a \textit{density-dependant correction} \cite{baskaran_self-regulation_2012} to the last term of Eq.\,(\ref{Eq:density}). 

The first term of Eq.\,(\ref{Q_corr}) promotes density dependent ordering, which competes with motility-induced disordering coming from the fist term of Eq.\,(\ref{Eq:Q}); $\beta$ is a non-equilibrium Landau coefficient setting the magnitude of order in the bulk.
$\tilde\kappa_{\rho}\langle\rho\rangle$ contributes to the restoring elastic constant. As can be seen, this is the only term in our theory that is linearized around the mean density value, whereas in the most of hydrodynamic models almost all terms in Eq.\,(\ref{colonly}) are subjected to this procedure. We linearize this particular term for two reasons. Firstly, for the sake of simplicity: we want this term to represent one particular effect -- elasticity (or ``rigidity'' in terms of the material). Secondly, with this linearization it's simpler to interpret the term $\tilde\kappa_{\rho} \langle\rho\rangle \Delta Q_{ij}$ as stemming from a free energy, while the contribution $\tilde\kappa\Delta(\rho Q_{ij})$ could not be obtained from a free energy.
Finally, the last term of Eq.\,(\ref{Q_corr}) describes the non-equilibrium anchoring to the density interface \cite{Maryshev2020}.

We emphasize again that we are not linearizing $\tilde\nu_{\rho},\,\tilde\chi_{\rho}$, and $\tilde\omega^a$  - related terms around the mean density (the latter of which would simply disappear completely in that case).
Such higher-order terms are typically linearized (or ignored) in well-controlled closures in the vicinity of the isotropic/nematic transition (e.g., within Boltzmann–Ginzburg–Landau approach \cite{ginelli_large-scale_2010,Peshkov2012,ngo_competing_2012-2, ngo_large-scale_2014}).
However, our observations hint that this linearization procedure, widely used in the field of active nematics, may result in some physical processes not being accounted for by the resulting models, which in turn can leads to some phenomena (e.g., such as CTDs) escaping the researchers' gaze as well.

To obtain the equations of motion presented in the main text we simply combine \eqref{advdiv} and \eqref{colonly} and re-normalize density by the critical one $\phi=\rho/\rho_\text{n}$. The coefficients are also renamed accordingly: $\tilde \kappa_{\rho}\rightarrow\kappa_{\phi}$, etc.

As discussed in the main text, the hydrodynamic model allows to directly access the direction and magnitude of the anisotropic active flux $-\partial_j(\chi Q_{ij})$. To complement the illustration of this flux in Fig.~4(d) of the main text, we show in Fig.~\ref{sfig:hydro_plot_Q} a direct plot of this observable as recorded in the hydrodynamic model.

\setcounter{figure}{0}

\renewcommand{\thefigure}{S\arabic{figure}}
\section*{ }

\clearpage

\clearpage
\renewcommand{\thefigure}{S\arabic{figure}}
\begin{figure}
\centering
\includegraphics[width=0.5\linewidth]{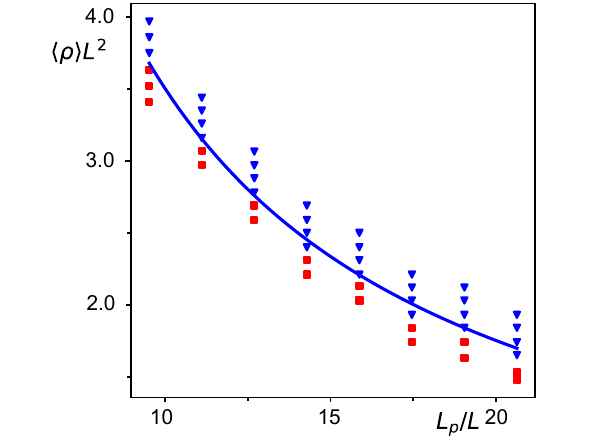}
\caption{
Phase space of nematic order. Agent-based simulations yielding nematic patterns are marked with blue triangles. Simulations exhibiting no order are shown as red squares. A fit of the functional form $f_{\rho}(L_p) = a/L_p$, where $a$ is the free fit parameter, to the ordered datapoints with the lowest density is shown in solid blue. (See Appendix for parameters.) 
}
\label{sfig:onset_nem}
\end{figure}

\clearpage

\begin{figure}
\centering
\includegraphics[width=0.6\linewidth]{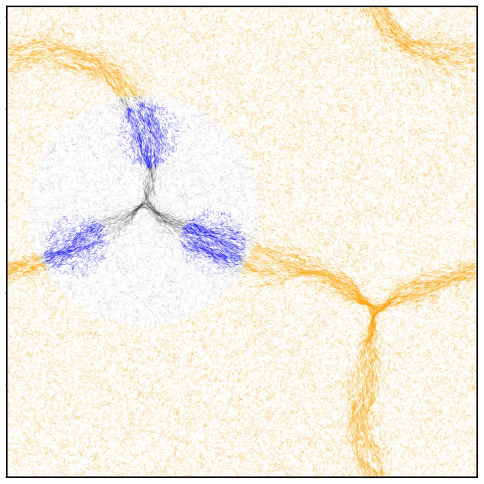}
\caption{
\textbf{Illustration of classification areas.} 
Representative spatial domain of an agent-based simulation containing 91217 polymers which exhibits two condensed topological defects (CTDs).
A circular classification area (regions where the polymers are colored blue) is placed over each nematic lane that emanates as an arm from one of the defects.
Polymers being further away from the defect than the blue classification areas are pooled into one large classification area (shown in orange). 
All black colored polymers do not belong to any classification area. 
(Parameters: ${L_p = 11L}$, simulation box size: $163L$; see Appendix for further parameters.)
}
\label{sfig:filter_regions}
\end{figure}

\clearpage

\begin{figure}
\centering
\includegraphics[width=0.8\linewidth]{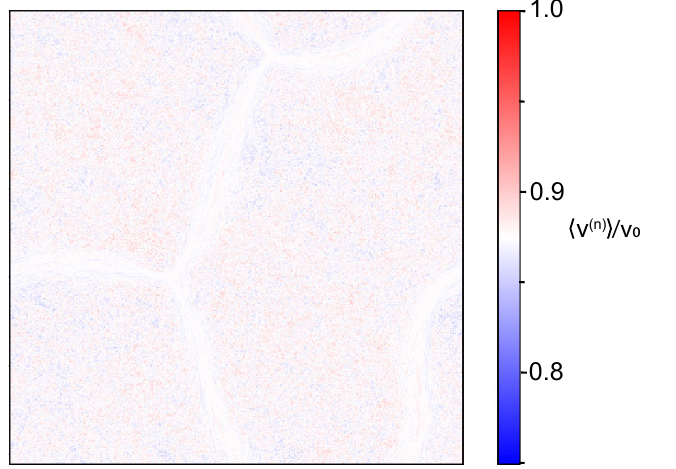}
\caption{
\textbf{Local averaged velocities.}
Local average of the free velocities $v^{(n)}$ of the agents in a system exhibiting nematic patterns. 
The values of the local averages, $\langle v^{(n)} \rangle$, do not deviate (up to random fluctuations) of the value obtained when averaging over all polymers inside the simulation ($\langle v^{(n)} \rangle_{\text{glob}}\,=\,0.875 v_0$, where $v_0$ is the maximal speed of a free polymer). 
Same data and parameters as used in Fig.~2(a) [the position of the patterns is still perceivable 
since fluctuations in the low density disordered background are less suppressed, due to the lower number of polymers over which is being averaged, compared with the high density nematic lanes.]
}
\label{sfig:local_avg_vel}
\end{figure}

\clearpage

\begin{figure}
\centering
\includegraphics[width=0.5\linewidth]{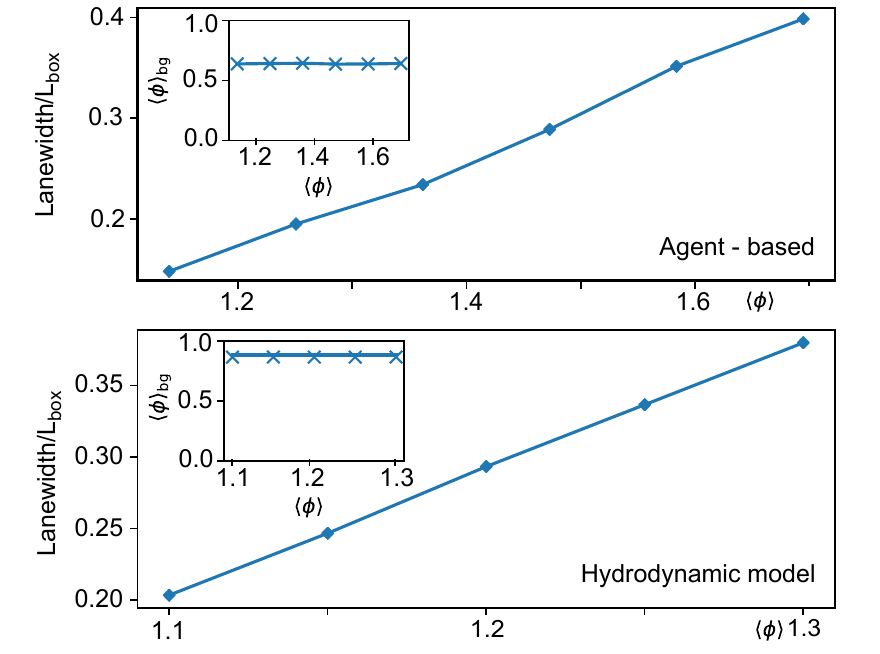}
\caption{
\textbf{Width of nematic lanes} for agent-based simulations (\textit{upper panel}) and for the hydrodynamic model (\textit{lower panel}). 
The width of stable nematic lanes grows with an increase of the global density $\langle \phi\rangle$ while the background density $\langle \phi\rangle _\text{bg}$ stays constant (inset).
Parameters: $L_p\,=\,20.6L$; see Appendix for further parameters.
}
\label{sfig:width_nem_lane}
\end{figure}

\clearpage

\begin{figure}
\centering
\includegraphics[width=0.5\linewidth]{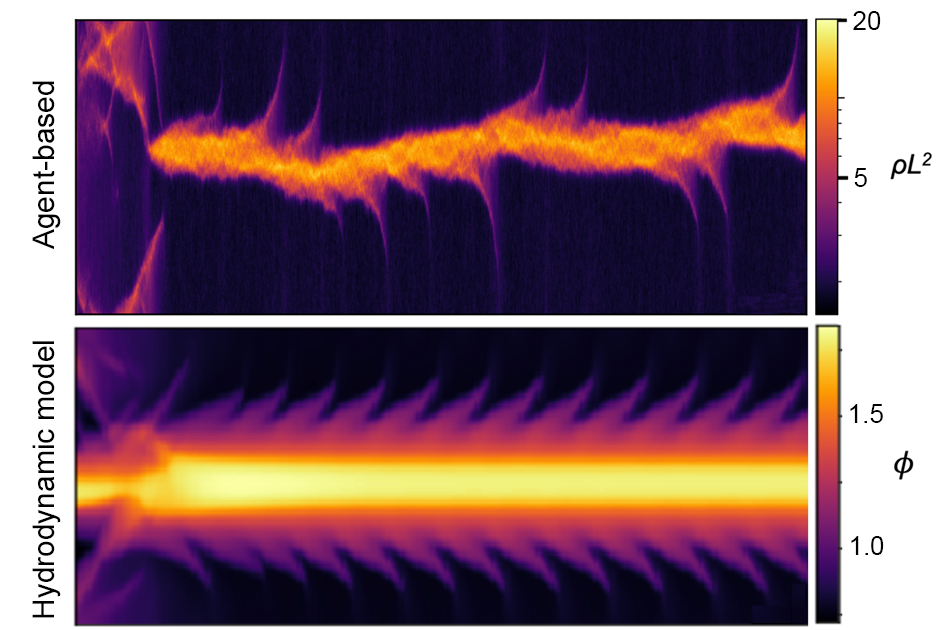}
\caption{
\textbf{Temporal evolution of FAEs} Illustration of a system exhibiting several filamentous arc ejections in sequence in agent based simulations (\textit{upper panel}) and hydrodynamic model (\textit{lower panel}). The density is projected/averaged along the long axis of the lane. The resulting 1-D slices are stacked into the shown Kymograph. Each FAE can be recognized by an extrusion from the lane. The slight bending of these extrusions towards a more vertical shape is a signature of the accelerated motion of the ejection.
}
\label{sfig:kymograph_fae}
\end{figure}

\clearpage

\begin{figure}
\centering
\includegraphics[width=0.5\linewidth]{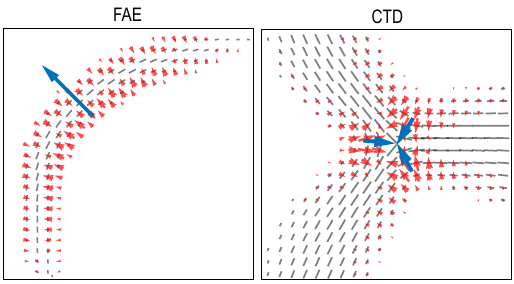}
\caption{
\textbf{Anisotropic active flux.}
Plot of $-\partial_j(\chi Q_{ij})$ in the hydrodynamic model. Grey segments represent nematic order and the red arrows corresponds to the magnitude and direction of the current. 
}
\label{sfig:hydro_plot_Q}
\end{figure}

\clearpage

\begin{figure}
\centering
\includegraphics[width=0.5\linewidth]{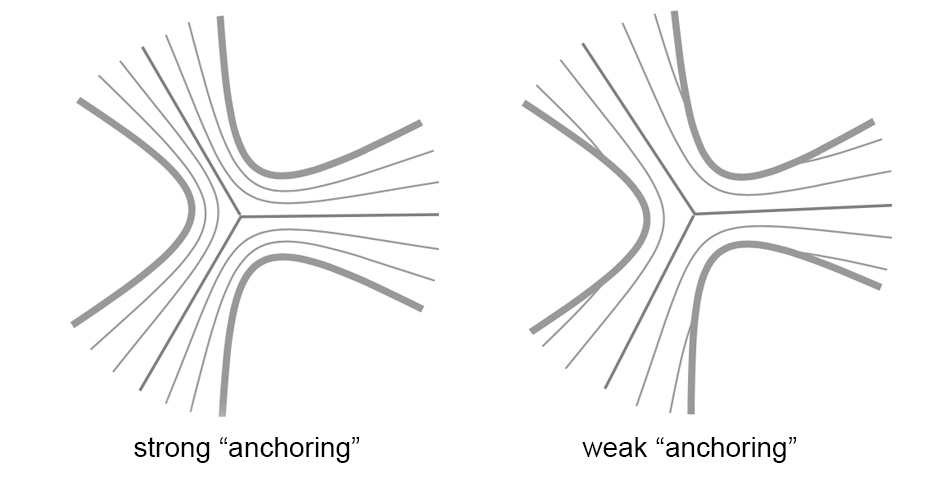}
\caption{
\textbf{Anchoring.}
Depictions of the nematic fieldlines inside a CTD with strong (large $|\omega^a|$) and weak (small $|\omega^a|$) bilinear anchoring term, respectively.
}
\label{sfig:anchoring_panels}
\end{figure}

\clearpage

\textbf{Movie S1}\\{\normalfont
Constantly undulating nematic lanes in an agent-based simulation.
(Parameters are: $\rho L^2{=}3.15$, $L_p{=}11.1$. Scale-bar: 15L. Density averaged over a time of 15.9 for better visibility.) }
\\ \\
\textbf{Movie S2}\\{\normalfont
Emergence of a multitude of condensed topological defects in agent-based simulations. Note that the lateral movement of lanes happens on long timescales. A single frame roughly corresponds to the time of 162 a straight moving particle with a velocity of $v_0$ needs to cross the whole system. (Parameters are: $\rho L^2{=}3.2$, $L_p{=}11.9$. Scale-bar: 15L. Density averaged over a time of 15.9 for better visibility.)}
\\ \\
\textbf{Movie S3}\\{\normalfont
Two condensed topological defects are formed simultaneously in an agent-based simulation. Due to continued undulation of the connecting nematic lanes the defects eventually disintegrate.
(Parameters are: $\rho L^2{=}3.47$, $L_p{=}11.1$. Scale-bar: 15L. Density averaged over a time of 3 for better visibility.)}
\\ \\
\textbf{Movie S4}\\{\normalfont
Several filamentous arc ejection develop in succession along a nematic lane in an agent-based simulation.
(Parameters are: $\rho L^2{=}2.7$, $L_p{=}14.3$. Scale-bar: 15L. Density averaged over a time of 15.9 for better visibility.)}
\\ \\
\textbf{Movie S5}\\{\normalfont
Straight and stable nematic lane in an agent-based simulation. 
(Parameters are: $\rho L^2{=}1.9$, $L_p{=}20.6$. Scale-bar: 15L. Density averaged over a time of 15.9 for better visibility.)}
\\ \\
\textbf{Movie S6}\\{\normalfont
Details of a flux in an agent-based simulation from one arm of a condensed topological defect to the two others. The path that is taken by the polymer heads is traced out. Only trajectories that start in the upper left arm and eventually will go to either the lower or upper right arm are visible.
(Parameters are: $\rho L^2{=}3.5$, $L_p{=}11.1$.)}
\\ \\
\textbf{Movie S7}\\{\normalfont
Emergence of a multitude of condensed topological defects in a simulation of the hydrodynamic model. (Parameters are: $\beta=0.05,\,\kappa_{\phi}=0.2,\,\omega^a=-0.5, \,\chi_{\phi}=0.4,\,\nu_{\phi}=1,\,\lambda=1,\langle\phi\rangle=1.1$ )}
\\ \\
\textbf{Movie S8}\\{\normalfont
Several filamentous arc ejection develop in succession along a nematic lane in a simulation of the hydrodynamic model. (Parameters are: $\beta=0.05,\,\kappa_{\phi}=0.2,\,\omega^a=-0.5, \,\chi_{\phi}=0.4,\,\nu_{\phi}=1,\,\lambda=1.2,\langle\phi\rangle=1.1$)}
\\ \\
\textbf{Movie S9}\\{\normalfont
Straight and stable nematic lane in a simulation of the hydrodynamic model. 
(Parameters are: $\beta=0.05,\,\kappa_{\phi}=0.2,\,\omega^a=-0.5, \,\chi_{\phi}=0.4,\,\nu_{\phi}=1,\,\lambda=1.4,\langle\phi\rangle=1.1$)}
\\ \\
\textbf{Movie S10}\\{\normalfont
Three-beam symmetrical arrangement of sources of polar particles. The ensuing nematic currents eventually form a condensed topological defect.
(Parameters are: $\rho L^2{=}3.6$, $L_p{=}14.3$. Scale-bar: 15L. Density averaged over a time of 15.9 for better visibility.)}
\clearpage

\end{document}